\documentclass[a4paper,11pt]{article}
\pdfoutput=1 
\usepackage{jcappub}
\usepackage[normalem]{ulem}
\usepackage[utf8]{inputenc}
\usepackage{enumerate}
\usepackage{amsmath, amssymb, amsthm, graphicx, epsfig, fancyhdr, slashed}
\usepackage[mathscr]{euscript}
\let\oldcdot\cdot
\usepackage{breqn}
\let\cdot\oldcdot
\usepackage{tikz}
\usetikzlibrary{shapes.geometric, arrows}
\usepackage{subcaption} 
\usepackage{url}
\usepackage{color,soul} 
\usepackage{multirow}
\usepackage{comment}
\usepackage{tabularx}
\usepackage{mathtools}
\usepackage{bm}
\usepackage{colortbl}
\usepackage{hyperref}
\usepackage[all]{hypcap}
\usepackage[verbose]{placeins}
\hypersetup{ 
    colorlinks=true,
    linkcolor=blue,
    filecolor=red,      
    urlcolor=purple,
    citecolor=green,
}

\title{\sf Neutrino self-interactions in post-reionization era: Lyman-$\alpha$, 21-cm and cross-spectra}
\author[a,1]{Sourav Pal,\note{Corresponding author.}}
\author[a]{and Supratik Pal}

\emailAdd{soupal1729@gmail.com}
\emailAdd{supratik@isical.ac.in}

\affiliation[a]{Physics and Applied Mathematics Unit, Indian Statistical Institute,\\ 203, B.T. Road, Kolkata 700108, India}

\abstract{
Neutrino self-interactions delay the onset of free-streaming in the early universe, leaving distinct, scale-dependent signatures on the matter power spectrum. We investigate these signatures in post-reionization 21-cm intensity mapping and the Lyman-$\alpha$ (Ly$\alpha$) forest at redshifts $z \sim 2$--$3.5$, and forecast the constraints achievable with upcoming surveys using  Fisher matrix analysis. Modeling neutrino self-interactions through an effective four-fermion parameterization with coupling $G_{\rm eff}$, we compute modifications to the Ly$\alpha$ and 21-cm auto- and cross-power spectra for both strongly interacting (SI$_\nu$, $\log_{10}G_{\mathrm{eff}} = -1.77$) and moderately interacting (MI$_\nu$, $\log_{10}G_{\mathrm{eff}} = -5$) scenarios. We then combine these with forecasts for a representative next-generation cosmic microwave background (CMB) mission to evaluate the capabilities of SKA1-Mid and PUMA. We find that the Ly$\alpha$--21-cm cross-correlation provides a systematics-resilient probe of the interaction signal, and  decisively breaks the degeneracy between the primordial scalar power spectrum amplitude ($A_s$) and $G_{\rm eff}$ that limits CMB only analysis, particularly for the SI$_\nu$ mode. Furthermore, the CMB+PUMA combination emerges as the optimal survey configuration for both regimes, reaching 1$\sigma$ constraints of $\mathcal{O}(10^{-3})$ on $\sigma(\log_{10}G_{\rm eff})$ for the SI$_\nu$ mode and $\mathcal{O}(10^{-2})$ for the MI$_\nu$ mode. Compared to the CMB-only baseline, this represents an improvement of approximately one order of magnitude for the SI$_\nu$ mode, and nearly two orders of magnitude for the MI$_\nu$ mode. We show that this conclusion holds uniformly over the full range of coupling strengths from $\log_{10}G_{\rm eff} = -6$ to $-1.77$. 
}

\begin{document}
\maketitle
\flushbottom

\section{Introduction}
\label{sec:introduction}
Neutrino oscillation experiments~\cite{Super-Kamiokande:1998kpq, Super-Kamiokande:2001bfk,
Super-Kamiokande:2002ujc, Super-Kamiokande:2005mbp, SNO:2002tuh, SNO:2002hgz,KamLAND:2004mhv,2003NuPhS:118146D} have firmly established that at least two of the three flavors of neutrinos are massive, providing the first confirmed evidence of physics beyond the Standard Model (BSM). However, terrestrial oscillation experiments can only measure mass-squared differences, leaving the absolute mass scale and potential non-standard interactions unconstrained~\cite{MINOS:2020llm,IceCubeCollaboration:2023wtb, NOvA:2023uxq}. Cosmological observations have the potential to fill this critical gap by providing non-trivial information from different epochs during the evolution history of the Universe. For the standard cosmological model, combining Cosmic Microwave Background (CMB) data from Planck and the Atacama Cosmology Telescope (ACT) with Baryon Acoustic Oscillation (BAO) measurements from the Dark Energy Spectroscopic Instrument (DESI) places stringent bounds on the sum of neutrino masses, restricting it to $\sum m_\nu < 0.072~\mathrm{eV}$ via the suppression of the power spectrum at non-linear scales~\cite{Planck:2018vyg, AtacamaCosmologyTelescope:2025nti,DESI:2024mwx, Allali:2024aiv, DESI:2025zgx, Elbers:2025vlz}. Along with putting formidable constraints to the commonly known neutrino parameters like $\sum m_\nu$ and the effective number of relativistic species $N_{\rm eff}$, cosmological data are highly sensitive to the free-streaming nature of neutrinos, offering a unique window into BSM neutrino physics that is fundamentally inaccessible to laboratory probes.

In the Standard Model, neutrinos decouple at temperatures $T \sim 1.5~\mathrm{MeV}$ and subsequently free-stream as collisionless particles. This behavior develops an anisotropic stress that shifts CMB acoustic peaks and suppresses matter clustering below the free-streaming horizon~\cite{Bashinsky:2003tk, Dolgov:2002wy, Lesgourgues:2006nd,Agarwal:2010mt, Blas:2014hya,Hu:1997mj, Saito:2008bp, Wong:2008ws,Levi:2016tlf, Pal:2023dcs, Nascimento:2023psl}. However, non-standard neutrino self-interactions can dramatically alter this picture. Typically parameterized by an effective four-fermion operator with a Fermi-like coupling constant $G_{\mathrm{eff}} = g_\nu^2 / m_\phi^2$ (arising from a massive scalar mediator $\phi$), the resulting self-interaction rate scales as $\Gamma_\nu \propto G_{\mathrm{eff}}^2 T_\nu^5$. If these interactions are sufficiently strong, they delay the onset of free-streaming until $\Gamma_\nu \sim H$, suppressing the anisotropic stress. Consequently, matter perturbation growth is enhanced on scales that enter the horizon while neutrinos remain tightly coupled, leaving a distinctive, scale-dependent imprint on cosmological observations~\cite{Cyr-Racine:2013jua, Oldengott:2017fhy, Kreisch:2019yzn,RoyChoudhury:2020dmd, Das:2020xke, Brinckmann:2020bcn, Das:2023npl,Pal:2024yom,Camarena:2023cku}.

Searches for these interaction signatures in CMB anisotropies have revealed a persistent and intriguing bimodality in the posterior of the self-coupling $G_{\mathrm{eff}}$~\cite{Cyr-Racine:2013jua,Kreisch:2019yzn, Das:2020xke,Das:2023npl}. Current data support both a moderately interacting (MI$_\nu$) mode, which largely resembles the Standard Model, and a strongly interacting (SI$_\nu$) mode, where free-streaming is delayed until near matter-radiation equality~\cite{Cyr-Racine:2013jua,Kreisch:2019yzn, Das:2020xke,Das:2023npl}. 
This bimodality persists across the CMB missions, namely, Planck and ACT analyses~\cite{Planck:2018vyg, Das:2023npl}. 
Evidence for the SI$_\nu$ mode was further established when a consistent preference at $\log_{10}(G_{\mathrm{eff}}/\mathrm{MeV}^{-2}) \approx -1.3$ was found in the Baryon Oscillation Spectroscopic Survey (BOSS) DR12 full-shape galaxy power spectrum. This independent detection strongly suggests that the SI$_\nu$ signature is not merely a statistical anomaly within CMB datasets, a conclusion further corroborated by recent Ly$\alpha$ and LSS analyses~\cite{Camarena:2023cku, Poudou:2025qcx,He:2023oke}.  
Despite these subtle hints, usual joint analyses combining CMB temperature, polarization, and galaxy power spectra disfavor the SI$_\nu$ mode compared to $\Lambda\mathrm{CDM}$, with Bayes factor at odds of $7:10000$~\cite{Camarena:2024daj}. This happens primarily because of the fact that Planck polarization data tightly constrains the primordial amplitude $A_s$, effectively excluding the low values required to accommodate the SI$_\nu$ in cosmology.

Above facts highlight a fundamental observational barrier: suppressed at small scales by Silk damping~\cite{1968ApJ...151..459S}, stand-alone CMB measurements lack sufficient resolving power needed to  constrain neutrino interactions conclusively. Overcoming this requires observational strategies that probe the matter power spectrum at the mildly non-linear regime  directly. 
Neutrino self-interaction signatures are particularly pronounced at galactic and sub-galactic scales: the SI$_\nu$ mode predicts a $30\text{--}40\%$ suppression of matter clustering driven by reduced $A_s$ and $n_s$~\cite{Camarena:2023cku, Camarena:2024daj}, whereas the MI$_\nu$ mode predicts a $10\text{--}15\%$ enhancement at $k \sim 1\text{--}10~h\,\mathrm{Mpc}^{-1}$ due to modes entering the horizon near neutrino self-decoupling~\cite{Camarena:2023cku, He:2023oke}. These features lie precisely in the regime accessible to complementary post-reionization probes, most notably 21-cm intensity mapping and the Ly$\alpha$ forest.

As individual probes, both 21-cm and Ly$\alpha$ observations hold immense potential to constrain cosmological parameters. The 21-cm hyperfine transition of neutral hydrogen provides a three-dimensional, wide-field map of the matter distribution in the post-reionization intergalactic medium at $z \lesssim 6$~\cite{Pritchard:2008da,Bull:2014rha,Liu:2019jiy,Bharadwaj:2000av,Camera:2015yqa,Villaescusa-Navarro:2014cma}. The integrated 21-cm emission from large cosmological volumes is measured without resolving individual galaxies, enabling efficient wide-field surveys of the matter power spectrum across a broad range of scales and redshifts~\cite{Modi:2019hnu,Sarkar:2016lvb,Libanore:2022ntl,Pal:2026hkq,Chakraborty:2020zmx}. Upcoming facilities like the Square Kilometre Array (SKA)~\cite{2015aska.confE..86N,SKA:2018ckk,Weltman:2018zrl} and the proposed Packed Ultrawideband Mapping Array (PUMA)~\cite{CosmicVisions21cm:2018rfq,PUMA:2019jwd,Castorina:2020zhz} aim to map this signal across broad scales and redshifts. The theoretical framework for observing neutrino-induced phase shifts in the 21-cm power spectrum, specifically isolating the contributions from BAO and velocity acoustic oscillations (VAO), has recently been established for the cosmic dawn epoch~\cite{Montefalcone:2025mbg}.  Recent forecasts already show that the Hydrogen Epoch of Reionization Array (HERA)~\cite{DeBoer:2016tnn}, combined with representative future CMB mission like CMB-S4, can break  degeneracies among different parameters for moderate interaction strengths~\cite{Libanore:2025ack}. Complementing this wide-field view, the Ly$\alpha$ forest captures small-scale density fluctuations ($k \sim 0.1\text{--}10~h\,\mathrm{Mpc}^{-1}$) via resonant photon absorption by neutral hydrogen (HI) along quasar sightlines~\cite{McDonald:2006qs,BOSS:2013rpr,Arinyo-i-Prats:2015vqa,deBelsunce:2024knf}. Recent analyses in~\cite{He:2023oke,Poudou:2025qcx} of the Ly$\alpha$ power spectrum from BOSS DR14 and extended BOSS (eBOSS) quasars indicated a strong preference for a strongly interacting mode; however, a later analysis in~\cite{He:2025jwp} favors the $\text{MI}_\nu$ mode over $\text{SI}_\nu$.


Despite a certain amount of promise shown by these individual measurements, their full observational potential is yet to be explored in a methodical way. Along with Ly$\alpha$ and 21-cm auto-correlations, cross-correlation of these two observations may turn out to be a treasure trove in this direction~\cite{Carucci:2016yzq,CHIME:2023til,Montero-Camacho:2024xvf}. 
Because 21-cm intensity mapping relies on radio interferometry and the Ly$\alpha$ forest on optical spectroscopy, their respective instrumental noises and foreground contaminations are completely statistically independent~\cite{Sarkar:2019yea,Dash:2020yuq,Dash:2020yfq}. Consequently, the cross-power spectrum cleanly isolates the shared cosmological signal while heavily suppressing individual systematic effects. In the context of interacting neutrinos, this cross-correlation retains the distinctive, scale-dependent morphological imprint of delayed free-streaming. Crucially, this imprint differs qualitatively between the SI$_\nu$ and MI$_\nu$ regimes, providing a robust, direct mechanism to distinguish the two modes and break the $G_{\mathrm{eff}}$, $A_s$, and $n_s$ degeneracies that currently limit CMB-only analyses. Notably, these parameter degeneracies have been shown to persist even when galaxy full-shape power spectrum data is included in the analysis, as has been investigated by the present authors~\cite{Pal:2024yom} as well as by some others~\cite{Camarena:2023cku}. In this context, post-Epoch of Reionization (post-EoR) multi-tracer observables emerge as effective probes of the neutrino interaction model. Serving as a comprehensive follow-up to our earlier investigation using CMB and galaxy full-shape  data~\cite{Pal:2024yom}, the present study methodically explores this observational potential of the post-reionization era.

Motivated by these prospects, this work investigates the signatures of early-universe neutrino self-interactions  on post-reionization 21-cm and Ly$\alpha$ tracers, forecasting the constraints achievable with next-generation surveys. Modeling the interactions via the standard effective four-fermion parameterization~\cite{Cyr-Racine:2013jua,Kreisch:2019yzn}, we compute the resulting modifications to the auto- and cross-power spectra in the mildly nonlinear regime. We then perform comprehensive Fisher matrix forecasts for SKA1-Mid, PUMA, and a representative future CMB mission based on CMB-S4 specifications~\cite{CMB-S4:2016ple,2017arXiv170602464A}. A central result of our analysis is that the 21-cm--Ly$\alpha$ cross-correlation acts as an incredibly robust observable: it decisively breaks the parameter degeneracies that have obscured the nature of BSM neutrino interactions in the CMB only analyses. We demonstrate that for the strongly interacting mode, the inclusion of 21-cm intensity mapping data effectively constrains the interaction coupling strength, tightening marginalized limits by a factor of 12 relative to the CMB-only baseline.  Furthermore, the interaction signature for the moderately interacting mode is shown to be strictly confined to high wave-numbers ($k \gtrsim 1 \, h \, {\rm Mpc}^{-1}$), yielding no observable imprint on primary CMB anisotropies.  High-resolution radio interferometric surveys, particularly PUMA, are thus established as uniquely capable of probing this regime, recovering the underlying cosmological signal and achieving an  enhancement in precision by nearly two orders of magnitude relative to the CMB baseline.

The paper is organized as follows. In Sec.~\ref{sec:neutrinos_CPT}, we review the effective four-fermion model and the modified Boltzmann hierarchy governing the cosmological evolution of interacting neutrinos. In Sec.~\ref{sec:sinu_ly_21}, we detail the theoretical modeling of the Ly$\alpha$ and 21-cm auto-power spectra and their cross-correlation as biased tracers of the matter field. Sec.~\ref{sec:Noise-modeling} describes the instrumental noise models for DESI-like Ly$\alpha$ spectroscopy, SKA1-Mid, PUMA, and CMB-S4 as a representative future CMB mission. In Sec.~\ref{sec:snr-fisher}, We present SNR maps for both the SI$_\nu$ and MI$_\nu$ scenarios in Ly$\alpha$ auto spectrum, 21-cm auto spectrum and the cross spectrum with and without considering the presence of an realistic foreground modeling. Furthermore, we outline the Fisher matrix methodology and present our joint parameter constraints for both the SI$_\nu$ and MI$_\nu$ scenarios across the full coupling parameter space.
We summarize our conclusions in Sec.~\ref{sec:summary}. Noise modeling and Fisher matrix framework for future CMB mission have been described in Appendix~\ref{app:cmb_noise}. All Fisher contours derived from stand-alone radio telescope surveys, as well as joint CMB and radio telescope observations, are provided in the Appendix~\ref{app:9params}.

\section{Neutrino Self-Interactions in Cosmological Perturbations}
\label{sec:neutrinos_CPT}
\subsection{Self-Interacting Neutrino Model and Boltzmann Hierarchy}
\label{subsec:sinu_model}
The impact of delayed neutrino free-streaming is efficiently captured using an effective four-fermion description~\cite{Bialynicka-Birula:1964ddi,Kreisch:2019yzn,Cyr-Racine:2013jua,Camarena:2023cku,Camarena:2024daj,Choudhury:2021dsc,Barenboim:2019tux,RoyChoudhury:2022rva}, in which neutrinos interact universally through a Fermi-like effective coupling constant $G_{\mathrm{eff}}$, leading to an interaction rate:
\begin{equation}\label{eq:Gamma_nu}
\Gamma_{\nu} \equiv a G_{\mathrm{eff}}^2 T_{\nu}^5\, ,
\end{equation}
where $a$ is the  scale factor and $T_{\nu}$ is the background neutrino temperature. The $T_{\nu}^5$ scaling reflects the dimensional structure of the four-fermion operator arising from phase-space factors governing $2 \to 2$ neutrino scattering at finite temperature. This representation can be thought of as arising from a Yukawa-type interaction between neutrinos and a scalar mediator $\phi$ of mass $m_{\phi} \sim \mathcal{O}(\text{MeV})$, through the relation $G_{\mathrm{eff}} = g_{\nu}^2 / m_{\phi}^2$, where $g_{\nu}$ is the neutrino--scalar coupling strength. Since $T_{\nu} \propto a^{-1}$ as the Universe expands, the interaction rate decreases as $\Gamma_{\nu} \propto a^{-4}$, ensuring that neutrino self-decoupling is inevitable for any finite $G_{\mathrm{eff}}$. 

The self-decoupling epoch is defined by $\Gamma_{\nu} \sim H$, which yields a characteristic decoupling temperature $T_{\mathrm{dec}} \sim \left[H(T_{\mathrm{dec}})/G_{\mathrm{eff}}^2\right]^{1/5}$. For the SI$_{\nu}$ mode with $\log_{10}(G_{\mathrm{eff}}/\mathrm{MeV}^{-2}) \approx -1.3$ to $-1.8$, as preferred by CMB and galaxy clustering data~\cite{Das:2023npl,Camarena:2023cku,Camarena:2024daj}, this corresponds to self-decoupling near the matter--radiation equality, eventually delaying free streaming relative to the $\Lambda$CDM predictions, whereas for the MI$_\nu$ mode, relatively weaker coupling results in an earlier self-decoupling deep within the radiation-dominated epoch.  Throughout this work, $G_{\mathrm{eff}}$ is expressed in units of $\mathrm{MeV}^{-2}$, with the standard weak Fermi constant corresponding to $G_{\mathrm{eff}} \sim \mathcal{O}(10^{-11}) \, \mathrm{MeV}^{-2}$. Following previous analyses~\cite{Cyr-Racine:2013jua,He:2023oke,RoyChoudhury:2020dmd,Camarena:2023cku}, we classify the parameter space into two distinct modes:
\begin{equation}
-5.5 \leq \log_{10}\left(\frac{G_{\mathrm{eff}}}{\mathrm{MeV}^{-2}}\right) \leq -2.5 \qquad (\mathrm{MI}_{\nu}) 
\end{equation}
\begin{equation}
-2.5 < \log_{10}\left(\frac{G_{\mathrm{eff}}}{\mathrm{MeV}^{-2}}\right) \leq 0.5 \qquad (\mathrm{SI}_{\nu})
\end{equation}
where the MI$_{\nu}$ mode produces modifications to nonlinear scales $k \gtrsim 0.5 \, h \, \mathrm{Mpc}^{-1}$, while the SI$_{\nu}$ regime induces substantial scale-dependent modifications at both linear and mildly nonlinear scales. We note that while this effective description accurately captures the leading-order effects of delayed free-streaming on cosmological observables, flavor-universal interactions of this form face tight constraints from Big Bang Nucleosynthesis (BBN), supernova cooling, and CMB polarization data~\cite{Das:2020xke}, implying that viable ultraviolet completions likely require non-trivial flavor structures. Nevertheless, we adopt this parameterization as a practical route enabling direct comparison with the existing literature. 

The evolution of neutrino perturbations in the presence of self-interactions is governed by a modified Boltzmann hierarchy incorporating the $\nu\nu \to \nu\nu$ collision term. We expand the neutrino temperature fluctuation in Legendre polynomials as~\cite{Kreisch:2019yzn,Das:2020xke,Camarena:2023cku,Berryman:2022hds,Perez-Castro:2026muj,Whitford:2025dmq}:
\begin{equation}
\frac{\delta T_{\nu}}{T_{\nu}}(\mathbf{k}, p, \tau) = \frac{1}{4}\sum_{\ell=0}^{\infty}(-i)^\ell(2\ell+1)\nu_{\ell}(k,p,\tau)P_{\ell}(\mu)  \, ,
\end{equation}
where $p = |\mathbf{p}|$, $k = |\mathbf{k}|$, and $\mu$ is the cosine of the angle between $\mathbf{k}$ and $\mathbf{p}$. Under the thermal approximation, the collision term at first order for the $\nu\nu \to \nu\nu$ process is:
\begin{equation}
C_{\nu}[p] = \frac{G_{\mathrm{eff}}^2 T_{\nu}^6}{4} \frac{\partial \ln f^{(0)}_{\nu}}{\partial \ln p} \sum_{\ell=0}^{\infty}(-i)^\ell(2\ell+1)\nu_{\ell} P_{\ell}(\mu) \left[A\left(\frac{q}{T_{\nu,0}}\right) + B_{\ell}\left(\frac{q}{T_{\nu,0}}\right) - 2D_{\ell}\left(\frac{q}{T_{\nu,0}}\right)\right] \, ,
\end{equation}
where $f^{(0)}_{\nu}$ is the background Fermi--Dirac distribution, $q = ap$ is the comoving momentum, $T_{\nu,0}$ is the present-day neutrino temperature, and $A(x)$, $B_{\ell}(x)$, $D_{\ell}(x)$ encode the angular structure of the collision integrals arising from the matrix element $|\mathcal{M}|^2 = 2G_{\mathrm{eff}}^2(s^2 + t^2 + u^2)$. Here $s,\,t,$ and $u$ are standard Mandelstam variables (see Refs.~\cite{Kreisch:2019yzn,Cyr-Racine:2013jua,RoyChoudhury:2020dmd,Das:2020xke,Camarena:2023cku} for more details). 

Adopting the conformal Newtonian gauge, the Boltzmann hierarchy for massive neutrinos reads:
\begin{align}\label{eq:boltzmann_hierarchy}
\frac{\partial \nu_{\ell}}{\partial \tau} &= -\frac{kq}{\epsilon}\left(\frac{\ell+1}{2\ell+1}\nu_{\ell+1} - \frac{\ell}{2\ell+1}\nu_{\ell-1}\right) + 4\left(\dot{\phi}\delta_{\ell 0} + \frac{k}{3}\frac{\epsilon}{q}\psi\delta_{\ell 1}\right) \nonumber \\
&\quad - \Gamma_{\nu}\frac{f^{(0)}_{\nu}(T_{\nu,0})}{q}\left[A \left(\frac{q}{T_{\nu,0}}\right) + B_{\ell}\left(\frac{q}{T_{\nu,0}}\right) - 2D_{\ell}\left(\frac{q}{T_{\nu,0}}\right)\right]\nu_{\ell}\, ,
\end{align}
where $\phi$ and $\psi$ are the scalar metric perturbations, and the comoving energy is $\epsilon = \sqrt{q^2 + a^2 m_{\nu}^2}$, with $m_{\nu}$ the neutrino mass. The above hierarchy satisfies both energy and momentum conservation exactly, since the collision terms vanish for $\ell \in \{0, 1\}$ moments. 
When $\Gamma_{\nu} \gg H$, frequent scatterings damp all multipoles with $\ell \geq 2$, driving $\nu_{\ell \geq 2} \to 0$ so that neutrinos behave as a perfect fluid with vanishing anisotropic stress. This profoundly alters the evolution of the gravitational potentials $\phi$ and $\psi$ and thereby modifies the growth of both photon and dark matter perturbations in a scale-dependent manner. Eq.~\ref{eq:boltzmann_hierarchy} is implemented in a modified version\footnote{\href{https://github.com/davidcato/class-interacting-neutrinos-PT}{https://github.com/davidcato/class-interacting-neutrinos-PT}} of the Boltzmann solver CLASS~\cite{blas2011cosmic,lesgourgues2011cosmic}, with a tight-coupling approximation employed at early times when $\Gamma_{\nu} > 10^3 H$ to avoid numerical stiffness, matching smoothly onto the full Boltzmann evolution as self-decoupling proceeds.

\subsection{Effects on Cosmological Observables}
\label{subsec:cosmo_observables}
The delay in the onset of neutrino free-streaming, induced by non-standard self-interactions, imprints distinctive, scale-dependent signatures on both the CMB anisotropies and the matter power spectrum~\cite{Cyr-Racine:2013jua, Kreisch:2019yzn}. The underlying physical mechanism is best elucidated by classifying Fourier modes according to their horizon-entry epoch relative to the era of neutrino self-decoupling, defined by the condition $\Gamma_{\nu} \sim H$. Three qualitatively distinct regimes emerge, parameterized by the effective coupling strength $G_{\mathrm{eff}}$:

\begin{itemize}
    \item \textbf{$k_{h}^{\mathrm{tc}}$}: Modes that cross the Hubble horizon while neutrinos remain tightly coupled (prior to self-decoupling).
    \item \textbf{$k_{h}^{\mathrm{fs}}$}: Modes entering the horizon near the onset of free-streaming (coincident with self-decoupling).
    \item \textbf{$k_{h}$}: Large-scale modes crossing the horizon well after the self-decoupling epoch has concluded.
\end{itemize}

\vspace{0.2em}
\noindent \textbf{Tightly coupled modes ($k_{h}^{\mathrm{tc}}$):} These modes enter the horizon during the strongly interacting phase. Consequently, the collision term effectively suppresses all higher-order multipoles ($\ell \geq 2$), forcing the gravitational potentials to satisfy $\psi \approx \phi$. The absence of neutrino anisotropic stress enhances the initial amplitude of the metric potential $\psi$ at horizon crossing. However, this missing anisotropic stress also amplifies the oscillatory envelope of $\psi$, driving a slower decay of the gravitational potential. This ultimately results in a net suppression of dark matter perturbation growth on these scales relative to the standard $\Lambda$CDM paradigm (a beautiful demonstration is provided in~\cite{Cyr-Racine:2013jua,Camarena:2023cku}).

\vspace{0.2em}
\noindent \textbf{Free-streaming onset modes ($k_{h}^{\mathrm{fs}}$):} Modes entering the horizon proximate to the self-decoupling epoch experience the initially enhanced $\psi$, but subsequently undergo the more rapid potential decay characteristic of free-streaming neutrinos once $\Gamma_{\nu} \lesssim H$. The dynamical interplay between these competing effects generates a net enhancement in dark matter clustering, manifesting as a localized excess in the matter power spectrum near $k_{h}^{\mathrm{fs}}$. Depending on the magnitude of $G_{\mathrm{eff}}$, this amplification can reach 10--15\%~\cite{Kreisch:2019yzn, Camarena:2023cku}.

\vspace{0.2em}
\noindent \textbf{Post-decoupling modes ($k_{h}$):} Large-scale modes that cross the horizon well after the conclusion of the self-interactions are effectively oblivious to the modified neutrino dynamics, recovering standard $\Lambda$CDM predictions.

\vspace{0.2em}

\begin{figure}[t]
\centering
\includegraphics[width=1.0\textwidth]{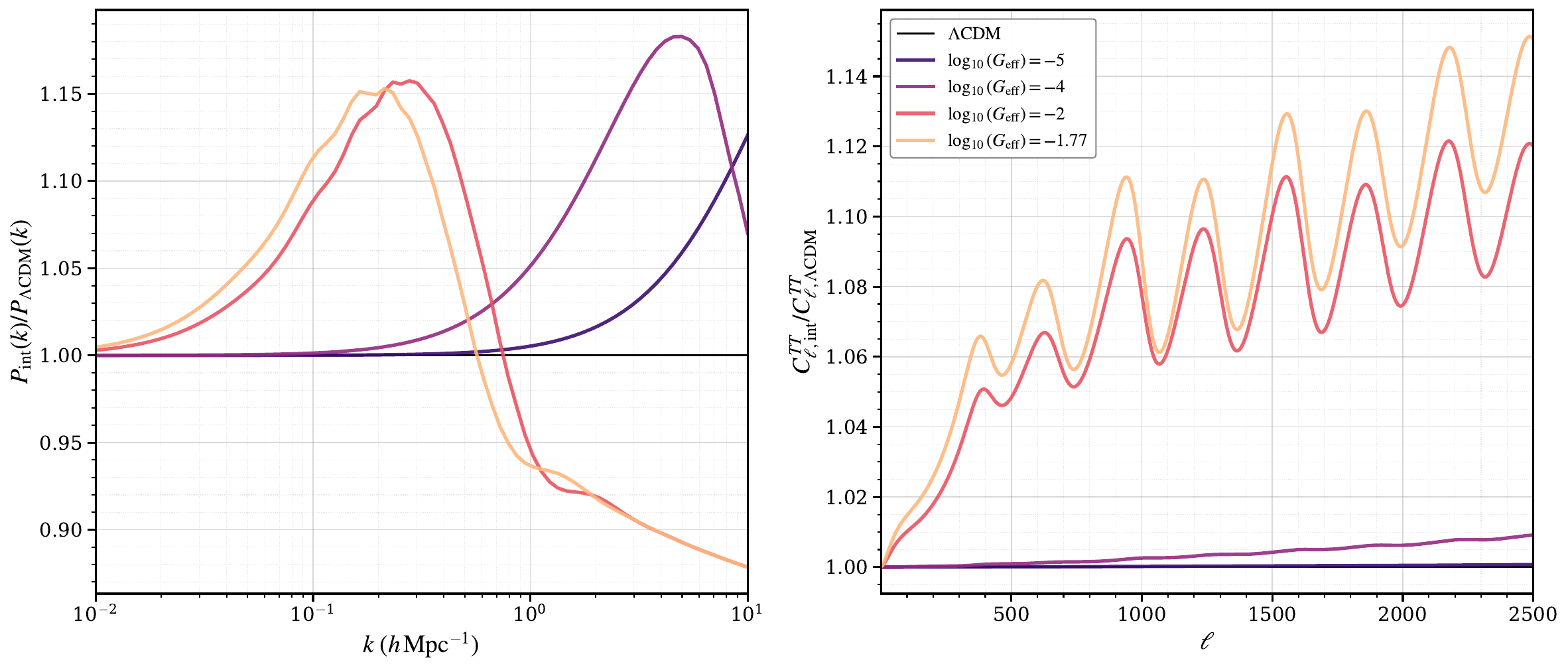}
\caption{Ratio of the self-interacting neutrino model to the standard $\Lambda$CDM baseline for the linear matter power spectrum (\textit{left panel}) and the CMB temperature power spectrum (\textit{right panel}). The curves highlight the scale-dependent impact of the effective coupling strength, $G_{\mathrm{eff}}$. Stronger couplings ($\log_{10}(G_{\mathrm{eff}}/\mathrm{MeV}^{-2}) = -1.77, -2$) induce large oscillatory deviations in the CMB and shift the matter power spectrum enhancement to linear/mildly non-linear scales ($k \sim 0.1$--$0.3 \, h \, \mathrm{Mpc}^{-1}$). Conversely, moderate couplings ($\log_{10}(G_{\mathrm{eff}}/\mathrm{MeV}^{-2}) = -4, -5$) leave the CMB spectrum essentially unperturbed while generating a distinct $15$--$18\%$ enhancement at small scales ($k \gtrsim 1 \, h \, \mathrm{Mpc}^{-1}$).}
\label{fig:ratio_ps}
\end{figure}
\subsubsection{Effects on the CMB power spectra}
In the standard $\Lambda$CDM paradigm, free-streaming neutrinos propagate supersonically following weak-interaction decoupling. This kinematics induces a gravitational drag on the photon–baryon plasma, generating a phase shift in the CMB acoustic peaks toward lower multipoles, $\ell$, a reduction in peak amplitude~\cite{Bashinsky:2003tk}.  This phase-shift, analytically established as a unique signature of free-streaming radiation under adiabatic initial conditions~\cite{Baumann:2015rya}, has been robustly detected in contemporary CMB observations~\cite{Follin:2015hya}. However, if neutrino free-streaming is delayed via the collisional suppression of higher-order multipoles ($\nu_{\ell \geq 2}$) during the tightly coupled phase, this gravitational drag is effectively nullified for modes entering the horizon prior to neutrino self-decoupling. Consequently, this dynamic results in amplified acoustic peak amplitudes and a compensatory phase shift toward higher $\ell$~\cite{Cyr-Racine:2013jua}. 
Recently, this compensatory shift has been utilized to construct template-based pipelines, allowing temperature-dependent interactions (including the four-fermion $G_{\rm eff}$ parameterization) to be directly mapped onto CMB observables to derive constraints from current Planck, ACT, and South Pole Telescope (SPT) data~\cite{Montefalcone:2025unv,Montefalcone:2025ibh}.

Phenomenologically, this effect manifests as a diverging oscillatory envelope for stronger coupling regimes ($\log_{10}(G_{\mathrm{eff}}/\mathrm{MeV}^{-2}) \geq -2$), as illustrated in the right panel of Fig.~\ref{fig:ratio_ps}. Additionally, the delayed free-streaming induces a fractional reduction in the photon sound horizon scale, a feature that can theoretically accommodate higher inferred values of the Hubble constant, $H_0$~\cite{Kreisch:2019yzn}. Conversely, for weaker couplings characteristic of the MI$_\nu$ regime ($\log_{10}(G_{\mathrm{eff}}/\mathrm{MeV}^{-2}) \leq -4$), the resulting CMB temperature spectra remain virtually indistinguishable from the baseline $\Lambda$CDM predictions. Despite its potential to alleviate the $H_0$ tension, high-precision CMB polarization data severely restrict this theoretical parameter space. Specifically, the strongly interacting mode necessitates anomalously low values for the primordial scalar amplitude, $\ln(10^{10}A_s)$. Such suppressed values are heavily disfavored by Planck limits ($10^9 A_s e^{-2\tau} \lesssim 1.85$). This stringent parameter requirement constitutes the primary observational tension currently challenging the viability of the SI$_\nu$ model~\cite{Planck:2018vyg, Camarena:2024daj}.

\subsubsection{Effects on the matter power spectrum}
The SI$_{\nu}$ mode, constrained by the necessary reductions in $\ln(10^{10}A_{s})$ and the scalar spectral index $n_{s}$, yields a matter power spectrum that is suppressed by over 30\% at galactic scales ($k \sim 1 \, h \, \mathrm{Mpc}^{-1}$) and by more than 40\% at sub-galactic scales ($k \gtrsim 10 \, h \, \mathrm{Mpc}^{-1}$), while retaining a marginal enhancement bump near $k \approx 0.1 \, h \, \mathrm{Mpc}^{-1}$~\cite{Camarena:2023cku}. This overall morphological structure exhibits remarkable consistency across both CMB-based fits and independent galaxy power spectrum analyses, suggesting a self-consistent cosmological picture~\cite{Camarena:2023cku, Camarena:2024daj}. 

In stark contrast, the MI$_\nu$ mode predicts a 10--15\% enhancement in the matter power spectrum at intermediate scales ($k \sim 1\text{--}10 \, \mathrm{Mpc}^{-1}$) without necessitating any shifts in $\ln(10^{10}A_{\mathrm{s}})$ or $n_{\mathrm{s}}$~\cite{He:2023oke}. As shown in Fig.~\ref{fig:ratio_ps}, these moderate couplings completely decouple the small-scale matter enhancement from the CMB, evading the stringent Planck polarization constraints that plague the SI$_\nu$ model. This provides a qualitatively distinct and readily detectable signal for surveys targeting small-scale structure, notably post-reionization 21-cm and Ly$\alpha$ observations. Crucially, accurately modeling these scales requires a robust treatment of massive neutrinos. The finite neutrino mass ($\sum m_{\nu} = 0.06 \, eV$) introduces an independent scale-dependent suppression of approximately $8 \, \Omega_{\nu}/\Omega_{m}$ for $k \gtrsim k_{\mathrm{fs}}^{\nu}$~\cite{Lesgourgues:2006nd, Planck:2018vyg}, where $\Omega_{\nu}$ is the neutrino energy density. Disentangling the signatures of finite mass from those of self-interactions
small-scale tracers like 21-cm intensity mapping and the Ly$\alpha$ forest, which distinctly probe the high-k domain~\cite{He:2023oke, He:2025jwp,Parashari:2026dxo}.

\section{Modeling the Ly$\alpha$ and 21-cm Power Spectra}
\label{sec:sinu_ly_21}
In this section, we outline the theoretical modeling of the Ly$\alpha$ and 21-cm power spectra utilized in our forecast analysis. The presence of neutrino self-interactions in the early universe alters the neutrino decoupling and consequently induce scale-dependent modifications
to the matter power spectrum~\cite{Kreisch:2019yzn, Cyr-Racine:2013jua}. 
In the post-reionization epoch (\textit{i.e.} $z \lesssim 6$), these scale-dependent modifications can be robustly probed using the Ly$\alpha$ forest and 21-cm intensity mapping, modeled as biased tracers of the underlying dark matter field in redshift space.
\subsection{Ly$\alpha$ Power Spectrum}
\label{subsec:ly}

The 3D Ly$\alpha$ flux power spectrum is a biased tracer of the underlying total matter power spectrum. In the mildly non-linear regime, it can be modeled using the Kaiser approximation for linear redshift space distortions (RSD)~\cite{Kaiser:1987qv}, supplemented by a phenomenological damping factor to account for the Finger-of-God (FoG) effect~\cite{Peebles:1980yev,Hikage:2015wfa,BaleatoLizancos:2025wdg}. 

The FoG effect arises from the random, non-linear peculiar velocity dispersion of intergalactic medium gas inside virialized structures. This thermal and virial motion kinematically smears the clustering signal along the line of sight, leading to a suppression of power at small scales~\cite{Jackson:1971sky}. As neutrino self-interactions predominantly modify the power spectrum at higher wavenumbers ($k \gtrsim 0.1 \, h \, \mathrm{Mpc}^{-1}$), accurately modeling the FoG damping is crucial to avoid degeneracies between the cosmological signal and non-linear astrophysical velocity effects. Incorporating both the linear RSD and the non-linear FoG damping, the 3D flux power spectrum is expressed as:
\begin{equation}
    P_F^{\mathrm{3D}}(\boldsymbol{k},z) = b_F^2\left(1+\beta_F\mu^2\right)^2 P_{\mathrm{m}}(k,z) e^{-k^2 \mu^2 \sigma_{v,F}^2}\, ,
\end{equation}
where $\mu \equiv k_{\parallel}/k$, $\sigma_{v,F}$ characterizes the non-linear velocity dispersion along the line of sight, and we utilize the approximate fitting forms for the flux bias, $b_F$, the linear RSD parameter, $\beta_F$, and FoG damping parameter $\sigma_{v,F}$ from \cite{Arinyo-i-Prats:2015vqa}, alongside the redshift extension factor provided in \cite{BOSS:2013rpr}. This term will essentially represent the contribution from cosmic variance--the two-point correlator of the Ly$\alpha$ flux fluctuations $\langle\delta_F\delta_F^*\rangle$ in Fourier space in the covariance matrix. 

The total observed power spectrum, $P_F^{\mathrm{T}}(\boldsymbol{k},z)$, is decomposed into the sum of three distinct components~\cite{McDonald:2006qs}:
\begin{equation}
    P_F^{\mathrm{T}}(\boldsymbol{k},z) = P_F^{\mathrm{3D}}(\boldsymbol{k},z) + P_F^{\mathrm{1D}}(k_{\parallel},z)P_w^{\mathrm{2D}}(z) + P_N^{\mathrm{eff}}(z)\, .
\end{equation}
The second term encapsulates the aliasing noise originating from the sparse sampling of quasars, where $P_F^{\mathrm{1D}}(k_{\parallel},z)$ is the 1D flux power spectrum measured along the line of sight, and $P_w^{\mathrm{2D}}(z)$ accounts for the spatial weights in the transverse direction, computed according to the analytic pipeline in \cite{McDonald:2006qs}. The third term, $P_N^{\mathrm{eff}}(z)$, represents the effective spectrographic instrumental noise. We have discussed in detail about the noise modeling in Sec.~\ref{subsec:lyman-alpha-noise}.
\subsection{21-cm Power Spectrum}
\label{subsec:21}
In the post-EoR,  the 3D power spectrum of the 21-cm fluctuations traces the aggregate neutral hydrogen (HI) emission. Including the aforementioned Kaiser RSD and FoG effects, it is expressed as:
\begin{equation}
    P_{21}^{\mathrm{3D}}(\boldsymbol{k},z) = \bar{T}(z)^2 b_{\mathrm{HI}}^2(k,z) \left(1+\beta_{\mathrm{HI}}\mu^2\right)^2 P_{\mathrm{m}}(k,z) D_{\mathrm{FoG}}(k_{\parallel}, \sigma_{v,21})\, ,
\end{equation}
with the global 21-cm brightness temperature, $\bar{T}(z)$, given by:
\begin{equation}\label{eq:Tbar}
    \bar{T}(z) = 188\,\frac{h}{E(z)}\,\Omega_{\rm HI}(z)\,(1+z)^2\;\rm mK \, ,
\end{equation}
where $E(z) = H(z)/H_0$ and $\Omega_{\rm HI}(z) = 4\times 10^{-4}(1+z)^{0.6}$ is the HI density parameter. 

At mildly non-linear scales, the spatial distribution of HI is governed by a scale-dependent, complex bias. As established by~\cite{Sarkar:2016lvb}, this non-linear HI bias function can be analytically approximated as a mixed polynomial:
\begin{equation}
    b_{\mathrm{HI}}(k,z) = \sum_{m=0}^4\sum_{n=0}^2 c(m,n)k^mz^n\, .
\end{equation}
Regarding its domain of validity, the coefficients $c(m,n)$ derived from baseline hydrodynamical simulations~\cite{Sarkar:2016lvb} yield a $b_{\mathrm{HI}}(k,z)$ that is robustly scale-independent at linear scales ($k \lesssim 0.1 \, \mathrm{Mpc}^{-1}$), but exhibits significant complex stochasticity at highly non-linear scales. The underlying simulations assume a minimum halo mass of $10^9 M_{\odot}$, which accurately captures the HI distribution at $z \leq 3.5$ but may slightly underestimate low-mass halo contributions at higher redshifts. The best-fit values of the polynomial coefficients are listed below, as obtained by \cite{Sarkar:2016lvb}:
\begin{equation}
    c(m,n)\times10^{2}=\begin{pmatrix} $65.31$ & $25.19$ & $1.963$ \\ $-60.74$ & $18.56$ & $1.806$ \\ $33.54$ & $-17.38$ & $1.618$ \\ $-5.129$ & $3.247$ & $-0.3803$ \\ $0.2773$ & $-0.1899$ & $0.02435$ \end{pmatrix}\:,
\end{equation}
where the rows correspond to $0\leq m\leq 4$ and the columns correspond to $0\leq n\leq 2$, and the associated $1\sigma$ errors are $\lesssim10\%$. These fitting forms hold for $k\lesssim20\:h/\textrm{Mpc}$ across the redshift range $z\sim2-6$.

For the RSD, N-body simulations indicate that a Lorentzian damping profile, $D_{\mathrm{FoG}} = (1 + k_{\parallel}^2 \sigma_{v,21}^2 / 2)^{-1}$, provides a statistically superior fit compared to a standard Gaussian. The pairwise velocity dispersion, $\sigma_{v,21}$, must incorporate intra-halo kinematics to accurately capture FoG suppression at $z \lesssim 6$. Its redshift evolution is well-described by the empirical fitting function~\cite{Sarkar:2018gcb}:
\begin{equation}
    \sigma_{v,21}(z) = \sigma_{v,0} (1 + z)^{-m} \exp\left[ - \left( \frac{z}{z_p} \right)^2 \right]\, ,
\end{equation}
where $\sigma_{v,0}=11\,{\rm Mpc}$, $m=1.9$, and $z_p=11$ are fitting parameters dependent on the specific HI mass assignment scheme.

\subsection{Ly$\alpha$--21-cm Cross-Power Spectrum}
\label{subsec:cross}
Extracting meaningful cosmological information from high-redshift auto-spectra often requires marginalization over complex astrophysical biases and tracer-specific foregrounds. Cross-correlating the Ly$\alpha$ forest with 21-cm intensity mapping inherently mitigates both issues. Because these two signals are acquired through completely independent observational channels, their  systematics and primary foregrounds are statistically uncorrelated~\cite{Sarkar:2019yea, Dash:2020yuq,Sarkar:2018vyh,Pal:2016icc,2011MNRAS.410.1130G}.
Furthermore, because these two observables trace the underlying dark matter distribution through different physical environments and bias mechanisms, their cross-spectrum helps break degeneracies between astrophysical effects and the amplitude of matter fluctuations. This makes the cross-correlation an exceptionally clean probe for isolating the pristine cosmological signal.

The cross-spectrum of Ly$\alpha$ and 21-cm, in the post-EoR, is given by:
\begin{equation}\label{eq:cross_spectra}
    P_{21,F}(\boldsymbol{k},z) = \bar{T}(z) b_F b_{\mathrm{HI}} \left(1+\beta_F\mu^2\right) \left(1+\beta_{\mathrm{HI}}\mu^2\right) P_{\mathrm{m}}(k,z) e^{-k^2 \mu^2 {\sigma_{v,F}^2}}\, .
\end{equation}
Where, in absence of any FoG modeling for the cross-spectrum, we consider the FoG parameters to be the same as the one considered for Ly$\alpha$ auto-spectrum.

The ratio of the self-interacting neutrino model to the standard $\Lambda$CDM model for the cross-spectrum reveals distinct, scale-dependent enhancements. For models within the SI$_\nu$ regime, the cross-spectrum exhibits a pronounced clustering enhancement at $k \sim 0.2-1.0 \ h \, \mathrm{Mpc}^{-1}$, followed by a sharp suppression at highly non-linear scales. In contrast, the MI$_\nu$ regime broadens and suppresses this signature. Because the specific shape and scale of these features are exquisitely sensitive to $G_{\rm eff}$, the cross-spectrum serves as a powerful tool to break parameter degeneracies between neutrino interactions and standard cosmic structure growth.

\section{Survey Specifications and Noise Modeling}
\label{sec:Noise-modeling}
Accurate forecasting of the statistical power for upcoming surveys necessitates rigorous modeling of the noise contributions to each power spectrum measurement. In this work, we consider three primary observables: the Ly$\alpha$ forest auto-power spectrum, the 21-cm intensity mapping auto-power spectrum, and their cross-power spectrum.
Because each observable possesses a distinct noise budget driven by the unique physical and instrumental characteristics of its respective survey, we outline the adopted noise models below. Specifically, we detail the instrumental configurations assumed for the DESI-like spectroscopic survey, the PUMA interferometer, and the SKA1-Mid single-dish telescope, as well as the noise configuration for the representative future CMB mission like CMB-S4.

\subsection{Radio Interferometry: SKA1-Mid and PUMA}
\label{subsec:21cm-noise}
For the 21-cm auto-power spectrum, the total noise comprises both the receiver's thermal noise and the Poisson shot noise stemming from the discrete nature of the HI source distribution. The total noise power spectrum is~\cite{Bull:2014rha,Castorina:2016bfm,Karagiannis:2020dpq,Jolicoeur:2020eup,Aharonian:2013av,Bonaldi:2025gpa}
\begin{equation}
    N_{21}(k, \mu, z) = N_{\rm thermal}(k, \mu, z) + \bar{T}^2(z)\,P_{\rm shot}(z),
\end{equation}
where $\bar{T}(z)$ represents the sky-averaged HI brightness temperature defined in Eq.~\ref{eq:Tbar} and $P_{\rm shot}(z)$ is the Poisson shot noise in $({\rm Mpc}/h)^3$. 
Both $P_{\rm shot}(z)$ and the HI bias $b_{\rm HI}(z, k)$ are taken from the tabulation ~\cite{Castorina:2016bfm} and interpolated as smooth functions of redshift. The spatial structure of the noise, however, relies on whether the instrument functions in an interferometer or single-dish capacity. We address these two operational modes separately below.

\subsubsection{PUMA (Interferometer Mode)}
The proposed design for PUMA utilizes a densely packed hexagonal grid of dishes, operating across a frequency band of 200–1100 MHz. This configuration allows PUMA to map the Universe over a redshift window of approximately $0.3<z<6$~\cite{CosmicVisions21cm:2018rfq}.
In interferometer mode, the instrument records the complex visibilities by cross-correlating pairs of antenna elements. The resulting thermal noise in the reconstructed power spectrum is dictated by the array's baseline density distribution, which governs the number of independent measurements contributing to each transverse Fourier mode. For a baseline $\mathbf{u}$ in the image plane, the effective number density of baselines is $n(u)$, where $u = |\mathbf{u}|$. For a specific transverse wavenumber $k_\perp$, the corresponding baseline length is $u = k_\perp \chi(z)/(2\pi)$, where $\chi(z)$ is the comoving angular diameter distance. The thermal noise power spectrum in the interferometer mode is given by~\citep{Karagiannis:2020dpq,Jolicoeur:2020eup}
\begin{equation}
    N_{\rm thermal}^{\rm IF}(k_\perp, z) = T_{\rm sys}^2(z)\,\frac{\chi^2(z)\,y(z)}{t_{\rm int}} \cdot \frac{\lambda^4(z)}{A_e^2} \cdot \frac{1}{2\,n(u, z)} \cdot \frac{S_{\rm area}}{\theta_{\rm FOV}(z)^2},
\end{equation}
where $\lambda(z) = 0.21(1+z)\,\rm m$ is the observed wavelength of the 21-cm line, $A_e = (\pi/4)D^2\eta$ is the effective collecting area of a single dish of physical diameter $D$ and aperture efficiency $\eta$, $t_{\rm int}$ is the total on-sky integration time, $S_{\rm area}$ is the survey solid angle, and $\theta_{\rm FOV}(z) = 1.22 \,\lambda(z)/D_{\rm Dish}$ is the primary beam's field of view. The cosmological conversion factor $y(z) = c(1+z)^2/(\nu_{21}H(z))$ translates a frequency bandwidth into a comoving line-of-sight depth. Additionally, the factor of 2 in the denominator accounts for the two polarization channels measured by each antenna element.

The baseline density $n(u, z)$ captures the specific array layout and determines the noise dependence on transverse scales. For a hexagonally close-packed array with $N_{\rm side}$ dishes per side, each of diameter $D$, the baseline density is well described by the empirical fitting function~\cite{CosmicVisions21cm:2018rfq,PUMA:2019jwd}
\begin{equation}
    n(u) = n_0\,\frac{a + b\,\tilde{u}}{1 + B\,\tilde{u}^{\,C}} \exp\!\left(-\tilde{u}^{\,D}\right), \qquad \tilde{u} = \frac{u}{N_{\rm side}\,D}, \qquad n_0 = \left(\frac{N_{\rm side}}{D}\right)^2,
\end{equation}
where the coefficients $(a, b, B, C, D)$ take different values for square-packed and hexagonally close-packed layouts. For PUMA's hexagonal packing, we adopt the fitted values $a = 0.5698$, $b = -0.5274$, $B = 0.8358$, $C = 1.6635$, $D = 7.3178$.

The system temperature entering the noise expression is
\begin{equation}
    T_{\rm sys}(z) = T_{\rm sky}(\nu) + T_{\rm scope},
\end{equation}
where the sky temperature is dominated by Galactic synchrotron emission, modeled by the power-law relation $T_{\rm sky}(\nu) = 25(\nu/400\,{\rm MHz})^{-2.75} + 2.75\,\rm K$. The receiver contribution, $T_{\rm scope} = T_{\rm ampl}/\eta_{\rm feed}^2 + T_{\rm ground}(1 - \eta_{\rm feed})/\eta_{\rm feed}$, incorporates both the amplifier noise temperature $T_{\rm ampl}$ and the ground spillover, which is defined by the feed efficiency $\eta_{\rm feed}$.

In interferometer mode, only a finite range of transverse wavenumbers is accessible, set by the minimum and maximum baselines of the array. Modes outside the range $k_{\min} = 0.01\,h\,{\rm Mpc}^{-1}$ and $k_{\max} = 2\,h\,{\rm Mpc}^{-1}$ are excluded from the Fisher analysis.
The instrumental specifications adopted for PUMA in this work are summarized in Table~\ref{tab:puma_specs}.

\begin{table}[h]
    \centering
    \renewcommand{\arraystretch}{1.3}
    \begin{tabular}{lc}
        \hline\hline
        Parameter & Value \\
        \hline
        Operating mode & Interferometer (hexagonal close-pack) \\
        Number of dishes per side, $N_{\rm side}$ & 256 \\
        Dish diameter, $D$ & 6\,m \\
        Aperture efficiency, $\eta$ & 0.7 \\
        Integration time, $t_{\rm int}$ & 5\,yr \\
        Sky fraction, $f_{\rm sky}$ & 0.5 \\
        Amplifier temperature, $T_{\rm ampl}$ & 50\,K \\
        Ground temperature, $T_{\rm ground}$ & 300\,K \\
        \hline\hline
    \end{tabular}
    \caption{Instrumental specifications for the PUMA interferometer configuration adopted in this work.}
    \label{tab:puma_specs}
\end{table}
\subsubsection{SKA1-Mid (Single-Dish Mode)}
The SKA1-Mid survey is intended to operate in single-dish (SD) mode across two primary frequency bands: Band 1 spanning $0.35<z<3.05$, and Band 2 spanning $0.10<z<0.49$ \cite{SKA:2018ckk,Jolicoeur:2020eup}. In single-dish mode, each dish operates independently as a total-power radiometer, mapping the sky by scanning across it. The individual maps from all dishes are subsequently co-added, reducing the thermal noise by a factor $\sqrt{N_{\rm dish}}$ relative to a single dish. Unlike the interferometer case, the thermal noise in single-dish mode is approximately isotropic in Fourier space to leading order, since there is no baseline density weighting by transverse scale. The thermal noise power spectrum is~\cite{Bull:2014rha,Karagiannis:2020dpq}
\begin{equation}
    N_{\rm thermal}^{\rm SD}(z) = \frac{T_{\rm sys}^2(z)\,\chi^2(z)\,y(z)\,S_{\rm area}}{2\,N_{\rm dish}\,t_{\rm int}},
\end{equation}
where $N_{\rm dish}$ is the total number of dishes, $t_{\rm int}$ is the total survey integration time, and the factor of 2 again accounts for dual-polarization measurements. The system temperature retains the same form as in the interferometer case, with $T_{\rm sys} = T_{\rm sky}(\nu) + T_{\rm rx}$, where $T_{\rm rx}$ is the receiver noise temperature of each dish.

The angular resolution of a single dish sets an upper limit on the accessible transverse wavenumbers. Modes with $k_\perp$ beyond the beam scale are exponentially suppressed in sensitivity. Rather than multiplying the noise expression by an explicit beam window function, we follow the approach of imposing sharp cuts on the accessible transverse scale range~\citep{Bull:2014rha}, where we set $k_{\rm min}=0.01 \, h\,{\rm Mpc}^{-1}$ and $k_{\rm max}= 2 \, h\, {\rm Mpc}^{-1}$ as like PUMA.
All modes with $k_\perp$ outside this range are excluded from the Fisher analysis. The SKA-Mid Band~1 specifications adopted in this work are given in Table~\ref{tab:ska_specs}.

\begin{table}[h]
    \centering
    \renewcommand{\arraystretch}{1.3}
    \begin{tabular}{lc}
        \hline\hline
        Parameter & Value \\
        \hline
        Operating mode & Single dish \\
        Number of dishes, $N_{\rm dish}$ & 197 \\
        Dish diameter, $D_{\rm dish}$ & 15\,m \\
        Survey area, $S_{\rm area}$ & 20\,000\,deg$^2$ \\
        Total integration time, $t_{\rm int}$ & 10\,000\,hr \\
        Receiver temperature, $T_{\rm rx}$ & 20\,K \\
        Aperture efficiency, $\eta$ & 1.0 \\
        \hline\hline
    \end{tabular}
    \caption{Instrumental specifications for the SKA-Mid Band~1 single-dish configuration adopted in this work.}
    \label{tab:ska_specs}
\end{table}

\subsection{Optical Surveys: DESI-like Ly\texorpdfstring{$\alpha$}{alpha} Noise}
\label{subsec:lyman-alpha-noise}

The noise in the Ly$\alpha$ forest three-dimensional power spectrum arises from two physically distinct sources. The first is the finite signal-to-noise ratio of individual quasar spectra, which sets a per-pixel noise floor that depends on the apparent magnitude of the background quasar, the spectrograph resolution, and the pixel width. The second is the discrete sampling of the flux field along lines of sight, which introduces an aliasing contribution whose amplitude is set by the one-dimensional power spectrum and the effective surface density of quasar lines of sight.

Following the optimal weighting framework of \cite{McDonald:2006qs,DESI:2024lzq,DESI:2024lzq,DESI:2023pir}, the total observed power in a given Fourier mode $(k, \mu)$ can be written as
\begin{equation}
    P_{\rm obs}(k, \mu) = P_F(k, \mu)\,W^2(k_\parallel) + P_{\rm alias}(k, \mu) + P_N(k, \mu),
\end{equation}
where $P_F(k,\mu)$ is the underlying Ly$\alpha$ flux power spectrum, $W(k_\parallel)$ is the spectral smoothing kernel, $P_{\rm alias}$ captures the aliasing contribution from the discrete line-of-sight distribution, and $P_N$ is the effective three-dimensional noise power arising from photon noise in the quasar spectra.

The smoothing kernel encodes both the pixelization of the spectrum and the finite spectral resolution of the spectrograph. For a pixel of width $\Delta v$ in km\,s$^{-1}$ and a Gaussian resolution kernel with one-dimensional width $\sigma_r$, the combined kernel takes the form
\begin{equation}
    W(k_\parallel) = \frac{\sin(k_\parallel \Delta v / 2)}{k_\parallel \Delta v / 2} \exp\!\left(-\frac{k_\parallel^2 \sigma_r^2}{2}\right),
\end{equation}
where the sinc factor arises from the finite pixel width and the Gaussian factor suppresses power on scales smaller than the resolution element. Both $\Delta v$ and $\sigma_r$ are specified in velocity units and depend on the observed wavelength and the spectrograph configuration.

The aliasing term arises because the three-dimensional Ly$\alpha$ flux field is only sampled along discrete quasar lines of sight. A mode with non-zero transverse wavenumber $k_\perp$ receives a spurious contribution from the one-dimensional power spectrum $P_{\rm 1D}(k_\parallel)$ along each line of sight. In the language of \cite{McDonald:2006qs}, this aliasing power is
\begin{equation}
    P_{\rm alias}(k, \mu) = \frac{\mathcal{I}_2}{\mathcal{I}_1^2\, L_F}\, P_{\rm 1D}(k_\parallel),
\end{equation}
where $L_F$ is the comoving length of the Ly$\alpha$ forest along the line of sight, and the integrals $\mathcal{I}_1$ and $\mathcal{I}_2$ over the quasar luminosity function are
\begin{align}
    \mathcal{I}_1 &= \int dm\, \frac{dn}{dm}(z)\, w(m), \\
    \mathcal{I}_2 &= \int dm\, \frac{dn}{dm}(z)\, w^2(m).
\end{align}
Here $dn/(dm)$(z) is the quasar number density per unit apparent magnitude, and $w(m)$ is the optimal weight assigned to quasars of apparent magnitude $m$. These weights balance the signal contribution against the noise variance and are determined iteratively, converging after a small number of iterations.

The effective three-dimensional noise power associated with photon noise in the spectra is
\begin{equation}
    P_N(k_\parallel) = \frac{\Delta v\,\mathcal{I}_3}{\mathcal{I}_1^2\, L_F},
\end{equation}
where the third luminosity-function integral is
\begin{equation}
    \mathcal{I}_3 = \int dm\, \frac{dn}{dm}(z)\, w^2(m)\, \sigma_N^2(m),
\end{equation}
and $\sigma_N^2(m)$ is the dimensionless pixel noise variance for a quasar of apparent magnitude $m$. This quantity is obtained from pre-computed signal-to-noise tables generated using the DESI instrument model. For a quasar of $r$-band magnitude $m$ observed at wavelength $\lambda_{\rm obs}$ with redshift $z_q$, the noise RMS per pixel of width $\Delta\lambda$ is
\begin{equation}
    \sigma_N(m, z_q, \lambda_{\rm obs}) = \frac{1}{(S/N)_{\rm pix}(m, z_q, \lambda_{\rm obs})},
\end{equation}
where the per-pixel signal-to-noise is related to the tabulated value via
\begin{equation}
    (S/N)_{\rm pix} = (S/N)_{\rm file} \cdot \sqrt{\Delta\lambda} \cdot \sqrt{\frac{N_{\rm exp}}{N_{\rm exp,\,file}}},
\end{equation}
with $(S/N)_{\rm file}$ computed for $N_{\rm exp,\,file} = 4$ exposures of $t_{\rm exp} = 4000\,\rm s$ each\footnote{The noise spectra for the Ly$\alpha$ auto and cross spectrum have been calculated using \href{https://github.com/igmhub/lyaforecast}{https://github.com/igmhub/lyaforecast}.}. If the quasar is brighter than the faintest magnitude tabulated in the files, the noise is evaluated at the file minimum; if it lies outside the tabulated redshift range, a large noise value is returned, effectively removing that quasar from the analysis. The total observed power entering the variance calculation is therefore
\begin{equation}
    P_{\rm tot}(k, \mu) = P_F(k, \mu)\,W^2(k_\parallel) + \frac{\mathcal{I}_2}{\mathcal{I}_1^2 L_F}\,P_{\rm 1D}(k_\parallel) + \frac{\Delta v\,\mathcal{I}_3}{\mathcal{I}_1^2 L_F}.
\end{equation}
Under the Gaussian approximation, the variance of the three-dimensional power spectrum estimator in a mode volume element $\Delta k\,\Delta\mu$ within a survey of comoving volume $V$ is
\begin{equation}
    \sigma^2_P(k, \mu) = \frac{2\,P_{\rm tot}^2(k, \mu)}{N_{\rm modes}}, \qquad N_{\rm modes} = \frac{V\,k^2\,\Delta k\,\Delta\mu}{2\pi^2}.
\end{equation}
The total Ly$\alpha$ power entering the cross-variance is expressed in comoving units via the Jacobian relation
\begin{equation}
    P_{\rm tot}^F(k,\mu)\bigl[({\rm Mpc}/h)^3\bigr] = P_{\rm tot}^F(k,\mu)\bigl[{\rm deg}^2\,{\rm km\,s}^{-1}\bigr] \cdot \frac{d_H^2(z)}{d_\parallel(z)},
\end{equation}
where $d_H(z)$ is the comoving angular diameter distance per degree in units of $h^{-1}$\,Mpc and $d_\parallel(z)$ is the comoving distance per unit velocity in $(h^{-1}\,{\rm Mpc})\,(\rm km\,s^{-1})^{-1}$. In the present study, we consider a next-generation DESI-like spectroscopic survey mission designed to achieve $\frac{dn}{dmdz}\sim100$, where $n$ is the number density of detected Ly$\alpha$ QSO samples, $m$ is the apparent magnitude. The expected number density of QSO samples is determined by the quasar luminosity function (QLF) \cite{Hopkins:2005ca,2013A&A...551A..29P,Yeche:2017upn,2023ApJ...949L..42M}.

\subsubsection{Ly\texorpdfstring{$\alpha$}{alpha}--21-cm  Cross-Power Spectrum Noise}

The cross-power spectrum between the Ly$\alpha$ forest and the 21-cm intensity field offers a complementary probe of large-scale structure, combining the high spectral resolution of optical spectroscopy with the wide-field survey capabilities and broad redshift reach of radio intensity mapping. Under the Gaussian approximation, the variance of the cross-power spectrum estimator is
\begin{equation}
    \sigma^2_{P_{\rm cross}}(k,\mu) = \frac{1}{2}[P_{\rm cross}^2(k,\mu) + P_{\rm tot}^F(k,\mu)\, \bigl[P_{21}(k,\mu) + N_{21}(k,\mu)\bigr]]\, ,
\end{equation}
where $P_{\rm cross}(k,\mu)$ is the Ly$\alpha$--21-cm cross-power spectrum, $P_{\rm tot}^F(k,\mu)$ is the total observed Ly$\alpha$ power including signal, aliasing, and spectroscopic noise, and $P_{21} + N_{21}$ is the total 21-cm power including both signal and thermal noise.

A key asymmetry in the cross-spectrum noise estimation is that only the Ly$\alpha$ spectrum is subject to spectral smoothing, since the 21-cm signal is not smoothed at the scales of interest by the radio telescope. Consequently, the cross-power is smoothed by a single power of the kernel $W(k_\parallel)$ rather than $W^2$ as in the Ly$\alpha$ auto-spectrum:
\begin{equation}
    P_{\rm cross}^{\rm obs}(k,\mu) = P_{\rm cross}(k,\mu)\,W(k_\parallel),
\end{equation}
which must be taken into account when computing the signal-to-noise ratio of the cross-correlation. The cross-power itself is defined in Eq.~\ref{eq:cross_spectra}.  
The combined variance of the cross-power estimator then follows directly from substituting the above into the Gaussian error expression and dividing by the number of independent Fourier modes in each bin.

\section{SNR Analysis and Fisher Forecast}
\label{sec:snr-fisher}
\subsection{Signal-to-Noise Ratio}
\label{subsec:snr}
To systematically evaluate the detectability of the delayed free-streaming signatures induced by neutrino self-interactions parameterized by $G_{\text{eff}}$, we compute the expected signal-to-noise ratio (SNR) for the mentioned three observables. Given a tracer power spectrum $P_i(k,z)$ and its associated variance $\sigma^2[P_i(k,z)]$, the SNR in a given redshift bin is expressed as:
$$
\text{SNR}_i^2 = \frac{V_{\text{survey}}(z_c) k^3 \epsilon \, d\mu}{4\pi^2} \frac{P_i(k,z)^2}{\sigma^2[P_i(k,z)]}
$$ \, ,
where $\epsilon \equiv dk/k$ is the logarithmic width of the $k$-bin, and $V_{\text{survey}}(z_c)$ is the comoving annular survey volume for a redshift bin centered at $z_c$. This volume is defined by the instrument's sky coverage and frequency bandwidth $\Delta\nu$ as:
$$
V_{\text{survey}}(z_c) = \frac{4\pi}{3} f_{\text{sky}} \left[ \chi(z_{\text{max}})^3 - \chi(z_{\text{min}})^3 \right] \, ,
$$
where $z_{\rm min}$ and $z_{\rm max}$ are the two redshift boundaries for the $z_c$-centered bin, determined by the frequency bandwidth of the instrument, $\Delta\nu$.
We anchor our forecast in realistic specifications for next-generation detectors. For the spectroscopic optical tracer, we utilize a DESI-like instrument mapping the Ly$\alpha$ forest over a sky area of 15,000~deg$^2$. For the radio emission, we consider two 21-cm intensity mapping arrays: SKA1-Mid, operating in single-dish mode over 20,000~deg$^2$ with a 10,000-hour total integration time, and PUMA, operating as a hexagonal close-pack interferometer over a sky fraction of $f_{\text{sky}} = 0.5$ with a 5-year integration time. Because the detectability of the Ly$\alpha - \text{21-cm}$ cross-correlation is fundamentally limited by their overlapping sky coverage, we restrict the cross-spectrum volume to a conservative overlap area of 2000~deg$^2$. Our purpose in this study is to demonstrate how these instrumental choices for next-generation detectors lead to considerably tight constraints on the neutrino interaction sector in the light of post-reionization observables. In particular, we analyze the SNR sensitivity as a function of both wavenumber and redshift, contrasting SI$_\nu$ (with $G_{\rm eff}=-1.77$) and  MI$_\nu$ (with $G_{\rm eff}=-5$) modes.

\begin{figure}[t]
\centering
\includegraphics[width=1.0\textwidth]{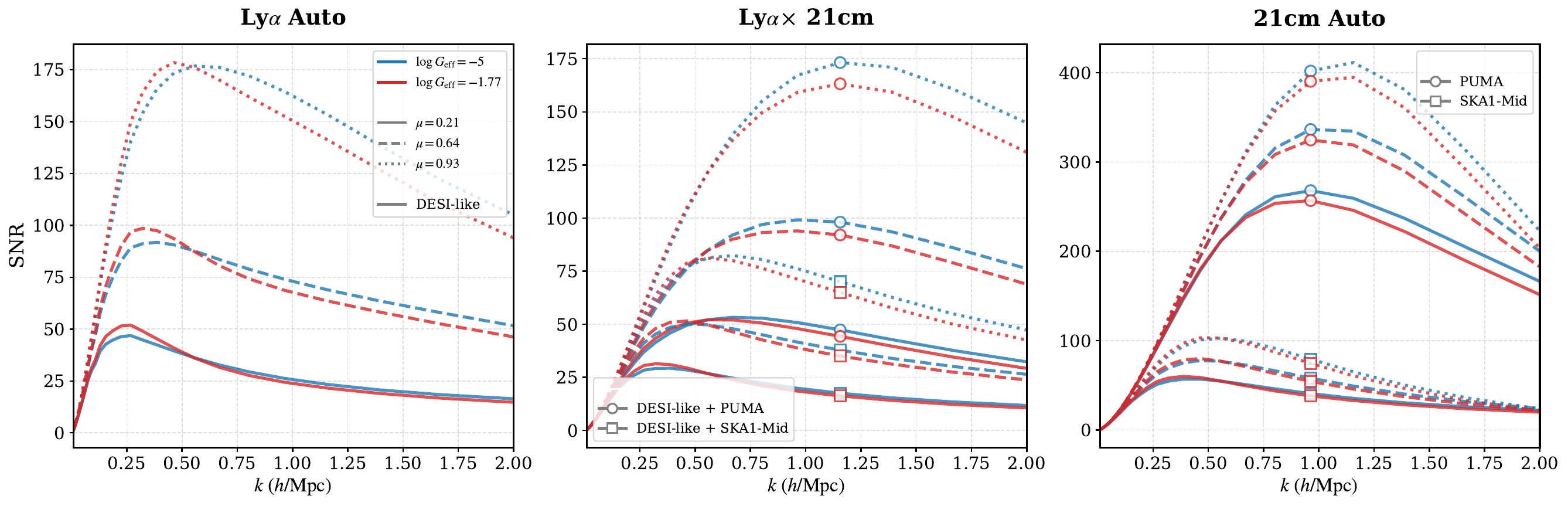}
\caption{Signal-to-noise ratio (SNR) for the Ly$\alpha$ auto, the Ly$\alpha - 21$-cm cross, and the 21-cm auto spectra shown as a function of scale, in the presence of neutrino self-interactions for the MI$_\nu$ ($\log_{10} G_{\mathrm{eff}} = -5$, blue) and SI$_\nu$ ($\log_{10} G_{\mathrm{eff}} = -1.77$, red) modes. The results are shown at a fixed redshift $z = 2.33$ for a few fixed values of the line-of-sight angle $\mu$. The middle and right panels show the cross- and auto-spectra for both the DESI-like+SKA1-Mid (dashed lines) and DESI-like+PUMA (dotted lines) configurations.}
\label{fig:snr_vs_k}
\end{figure}

Fig.~\ref{fig:snr_vs_k} illustrates the forecasted SNR as a function of k at a representative redshift of $z = 2.33$ for the Ly$\alpha$ auto-spectrum, the Ly$\alpha - 21$-cm cross-spectrum, and the 21-cm auto-spectrum. Modes more closely aligned with the line of sight yield a substantially higher SNR, highlighting a strong and universally observed dependence on the line-of-sight angle ($\mu$).
The observed amplification is largely attributable to linear redshift-space distortions; however, this enhancement is progressively attenuated at high k due to non-linear FoG effects. Assuming an optimistic, foreground-free scenario, the overall SNR values for the three signals exhibit the following trend\footnote{We note that the 21 cm auto-spectrum currently outperforms the Ly$\alpha$-21 cm cross-spectrum in SNR strictly due to our simplified, foreground-free noise modeling and broad sky coverage. Future data will be the true test of this feature; once realistic foregrounds are factored into the analysis, the current inequalities may no longer hold. The effects of realistic foreground modeling on the SNR have shown in Sec.~\ref{sec:foregrounds}.}:
\begin{equation}
    \textrm{SNR}_{{\rm Ly}\alpha -{\rm auto}}<\textrm{SNR}_{{{\rm Ly}\alpha}-21cm}<\textrm{SNR}_{\rm 21cm-auto}\:.
\end{equation}

As can be found from Fig.~\ref{fig:snr_vs_k}, the delayed free-streaming induced by neutrino self-interactions leads to scale-dependent modifications in the SNR for all three observables. The SI$_\nu$ mode with $\log_{10} G_{\mathrm{eff}} = -1.77$ (while other parameters are fixed to Planck18 baseline values~\cite{Planck:2018vyg}) typically dominates the SNR at intermediate scales ($k \sim 0.2$ to $0.5\,h\,\text{Mpc}^{-1}$) due to its characteristic enhancement bump, whereas at larger scales, the SNR curves converge to a common ($\Lambda$CDM) baseline as expected. In contrast, the MI$_\nu$ mode with $\log_{10} G_{\mathrm{eff}} = -5$ overtly surpasses the SI$_\nu$ mode in all three observables at smaller scales (typically $k \sim 0.5$ to $2\,h\,\text{Mpc}^{-1}$). This happens because the MI$_\nu$ mode mainly enhances small-scale power, whereas the SI$_\nu$ mode severely suppresses it. Therefore, precise measurements of high-$k$ modes are crucial for distinguishing between these two neutrino interaction regimes. 
Further, at $k \approx 1.0\,h\,\text{Mpc}^{-1}$ evaluated at a line-of-sight cosine $\mu=0.93$, PUMA provides an SNR approaching 400 for the 21-cm auto-spectrum, whereas SKA1-Mid gives SNR $\sim 100$.
Thus, the cross-power spectrum and the 21-cm auto-power spectrum underscore an instrumental sensitivity of the PUMA interferometric array at  a relatively higher level than SKA1-Mid.  

\begin{figure}[t]
\centering
\includegraphics[width=1.0\textwidth]{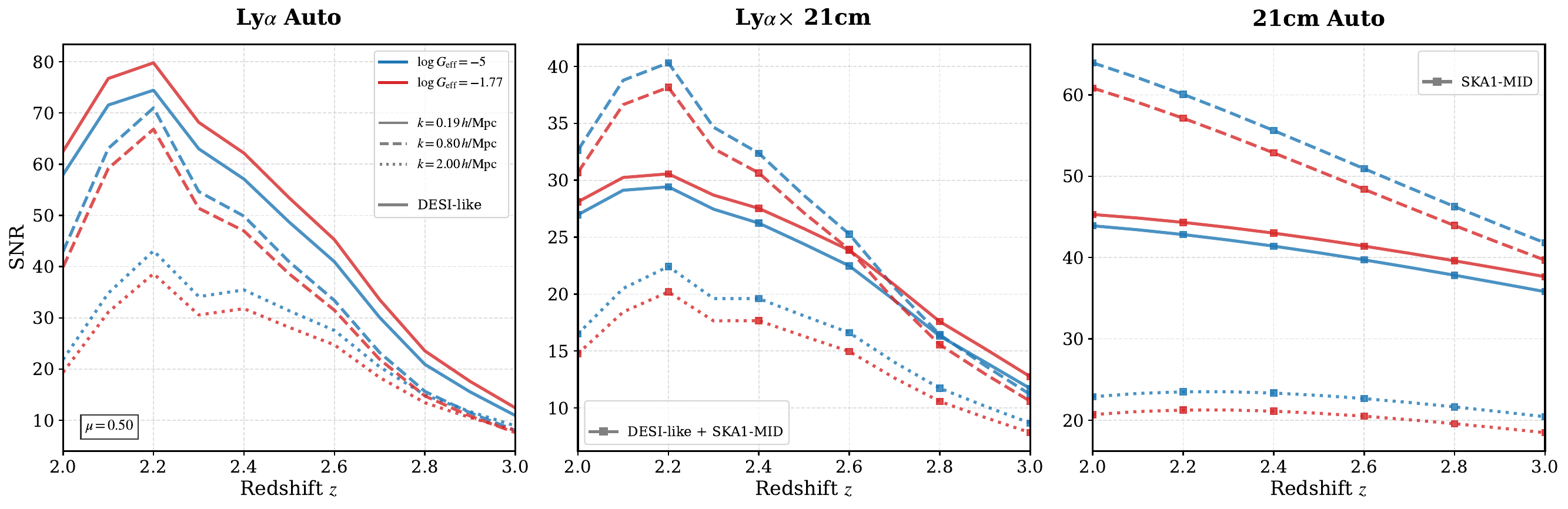}
\caption{Signal-to-noise ratio (SNR) for the Ly$\alpha$ auto, the Ly$\alpha - 21$-cm cross, and the 21-cm auto spectra shown as a function of redshift, in the presence of neutrino self-interactions for the MI$_\nu$ ($\log_{10} G_{\mathrm{eff}} = -5$, blue) and SI$_\nu$ ($\log_{10} G_{\mathrm{eff}} = -1.77$, red) modes. The different curves correspond to a few benchmark values of the scale $k \in \{0.19, 0.80, 2.00\} \, h/\mathrm{Mpc}$, and $\mu = 0.50$ has been fixed for the purpose of illustration. For the instruments, we assume the DESI-like spectroscopic detector (for Ly$\alpha$) and the SKA1-Mid configuration (for 21-cm).}
\label{fig:snr_vs_z_ska}
\end{figure}

\begin{figure}[t]
\centering
\includegraphics[width=1.0\textwidth]{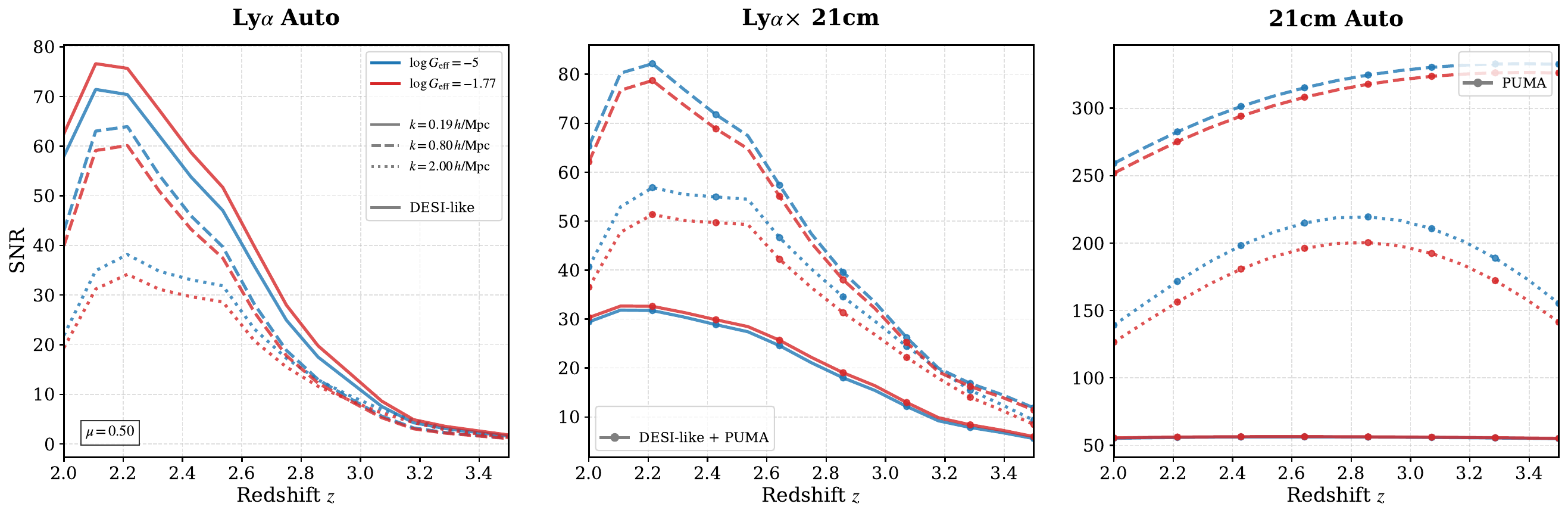}
\caption{Redshift evolution of the SNR at a fixed angle ($\mu=0.50$), analogous to Fig.~\ref{fig:snr_vs_z_ska}, but utilizing PUMA for the 21-cm observations. The panels show the DESI-like Ly$\alpha$ auto-spectrum (\textit{left panel}), the DESI-like+PUMA cross-spectrum (\textit{middle panel}), and PUMA 21-cm auto-spectrum (\textit{right panel}).}
\label{fig:snr_vs_z_puma}
\end{figure}

On the other hand, the variation of the SNR with redshift  reveals distinct observational windows for these surveys, as shown in Figs.~\ref{fig:snr_vs_z_ska} and \ref{fig:snr_vs_z_puma} for a fixed orientation of $\mu=0.50$. In both the figures, the SNR for both the Ly$\alpha$ auto and the Ly$\alpha-$ 21-cm  cross spectra are seen to peak across the intermediate redshift range 2.1 to 2.4, falling off towards higher values of $z$ due to increasingly sparse sampling of quasars and instrumental noise. 
The SNR for the 21-cm power spectrum, on the other hand, exhibits differing behaviors depending on the instrument. For SKA1-Mid (Fig.~\ref{fig:snr_vs_z_ska}), the SNR declines towards higher redshifts. However, for the interferometric sensitivity of PUMA (Fig.~\ref{fig:snr_vs_z_puma}), the SNR rises monotonically towards the higher end of the observable redshift range ($z \sim 2.8$ to $3.0$). However, the 21-cm auto-spectrum
is expected to be severely affected by atmospheric and extraterrestrial foregrounds, which may
reduce the optimistic SNR by several orders of magnitude~\cite{Wolz:2015sqa,Zuo:2022wra,Diao:2024dmu}. While the Ly$\alpha$ auto-spectrum avoids foreground contamination, its SNR is inherently much lower than that of the 21-cm auto-spectrum or the cross-spectrum. The Ly$\alpha-21$-cm  cross-correlation offers an ideal compromise, providing an intermediate SNR that is largely unaffected by foregrounds \cite{Carucci:2016yzq}. This stability makes the cross-spectrum a  unique late-time tool for probing neutrino self-interactions.

\subsection{Fisher Matrix Forecast Analysis}
\label{subsec:fisher}
In what follows, we employ the Fisher information matrix forecast formalism~\cite{Tegmark:1996bz,2009arXiv0906.0664H,Bharadwaj:2015vwa} to estimate the precision in future measurements of the interacting neutrino parameter space. Our analysis is based on a combination of the LSS tracer observables and CMB-S4, a representative future CMB mission.
 For a Gaussian likelihood $\mathcal{L}$, the Fisher matrix elements are given by the expectation value of the log-likelihood:
\begin{equation}
    F_{\alpha\beta} = -\left\langle \frac{\partial^2 \ln \mathcal{L}}{\partial \theta_\alpha \, \partial \theta_\beta} \right\rangle.
\end{equation}
In the present scenario, for a set of $n$ fiducial parameters encoded in the parameter vector $\boldsymbol{\Theta}$, the $(n\times n)$-dimensional Fisher matrix corresponding to the $i$-th observable can be expressed as
\begin{equation}
    F_{\alpha\beta}^{(i)}(\boldsymbol{\Theta})=\sum\limits_{m}^{\rm k-bins}\sum\limits_{n}^{\rm z-bins}\sum_{p}^{\mu{\rm -bins}}\:\dfrac{1}{\sigma^2\left[P_i\left(k_m,z_n,\mu_p\right)\right]}\dfrac{\partial P_i\left(k_m,z_n,\mu_p\right)}{\partial\Theta_\alpha}\dfrac{\partial P_i\left(k_m,z_n,\mu_p\right)}{\partial\Theta_\beta}\:,
\end{equation}
where $\alpha$ and $\beta$ run from $1$ to $n$. The inverse of the Fisher matrix then gives the covariance matrix, whose diagonal elements correspond to the projected $1\sigma$ errors for the parameters, and off-diagonal elements correspond to the correlation among different parameters.

To fully capture the physical degeneracies between the neutrino interaction strength and standard cosmological parameters, we take our full parameter vector $\boldsymbol{\theta}$ as:
\begin{equation}
    \boldsymbol{\theta} = \left\{ \omega_b, \omega_c, h, \tau, \ln(10^{10}A_s), n_s, \log_{10}\!\left(\frac{G_{\rm eff}}{{\rm MeV}^{-2}}\right), M_\nu, N_{\rm eff} \right\},
\end{equation}

Because the CMB and post-reionization LSS measurements probe independent volumes and linear/mildly non-linear regimes respectively, their datasets are statistically independent. The final forecasted parameter covariance matrix is therefore the inverse of the sum of the individual Fisher matrices:
\begin{equation}
    \mathbf{C}_{\alpha\beta} = \left( F^{\rm LSS} + F^{\rm CMB} \right)^{-1}_{\alpha\beta}.
\end{equation}
The marginalized $1\sigma$ uncertainty on any parameter $\theta_\alpha$, including our primary target $\log_{10}(G_{\rm eff}/{\rm MeV}^{-2})$, is simply given by $\sqrt{\mathbf{C}_{\alpha\alpha}}$. Fisher matrix for the CMB observables, $F^{\rm CMB}_{\alpha \beta}$ is detailed in Appendix~\ref{app:cmb_noise}.

\subsection{Results and discussions}
\label{subsec:results}

The fiducial cosmologies adopted for each interaction regime for the forecast analysis have been summarized in Table~\ref{tab:fiducial_params}. The SI$_\nu$ fiducial is fixed at $\log_{10}(G_{\rm eff}/{\rm MeV}^{-2}) = -1.77$, consistent with the phenomenologically preferred region identified in combined CMB and galaxy power spectrum analyses~\cite{Camarena:2023cku, Das:2023npl}. To remain consistent with CMB observations, this SI$_\nu$ configuration requires correspondingly lower values for the primordial amplitude ($10^9 A_s = 1.959$) and spectral index ($n_s = 0.9298$). The MI$_\nu$  fiducials, by contrast, are chosen from a value consistent with $\Lambda$CDM,
since the matter power spectrum enhancements in those regimes arise purely from the modified neutrino dynamics rather than from any shift in the primordial sector.

\begin{table}[h]
\centering
\renewcommand{\arraystretch}{1.3}
\begin{tabular}{lccc}
\hline\hline
Parameter & $\Lambda$CDM & SI$_\nu$ & MI$_\nu$ \\
\hline
$\log_{10}(G_{\rm eff}/{\rm MeV}^{-2})$ & --- & $-1.77$  & $-5$ \\
$h$ & $0.673$ & $0.670$  & $0.673$ \\
$\omega_b$ & $0.022329$ & $0.022355$  & $0.022329$ \\
$\omega_c$ & $0.118305$ & $0.116490$  & $0.118305$ \\
$\ln(10^{10}A_s)$ & $3.0412$ & $2.9750$  & $3.0412$ \\
$n_s$ & $0.9637$ & $0.9298$  & $0.9637$ \\
$N_{\rm eff}$ & $3.01$ & $2.82$ & $3.01$ \\
$M_\nu\,[\rm eV]$ & $0.06$ & $0.080$  & $0.06$ \\
$\tau$ & $0.052$ & $0.045$  & $0.053$ \\
\hline\hline
\end{tabular}
\caption{Fiducial cosmological parameters adopted for the Fisher matrix forecasts in each neutrino interaction scenario. The SI$_\nu$ fiducial follows the best-fit region preferred by combined CMB and galaxy power spectrum analyses~\cite{Camarena:2023cku, Das:2023npl}, which requires reduced values of $A_s$ and $n_s$ relative to $\Lambda$CDM. The MI$_\nu$ fiducial retain standard primordial parameters, since the matter power spectrum enhancement in these regimes arises purely from the modified neutrino dynamics.}
\label{tab:fiducial_params}
\end{table}
\subsubsection{SI$_\nu$ Mode}
\label{subsubsec:SI_results}
\begin{figure}[ht!]
    \centering
    \subfloat{\includegraphics[width=0.55\textwidth]{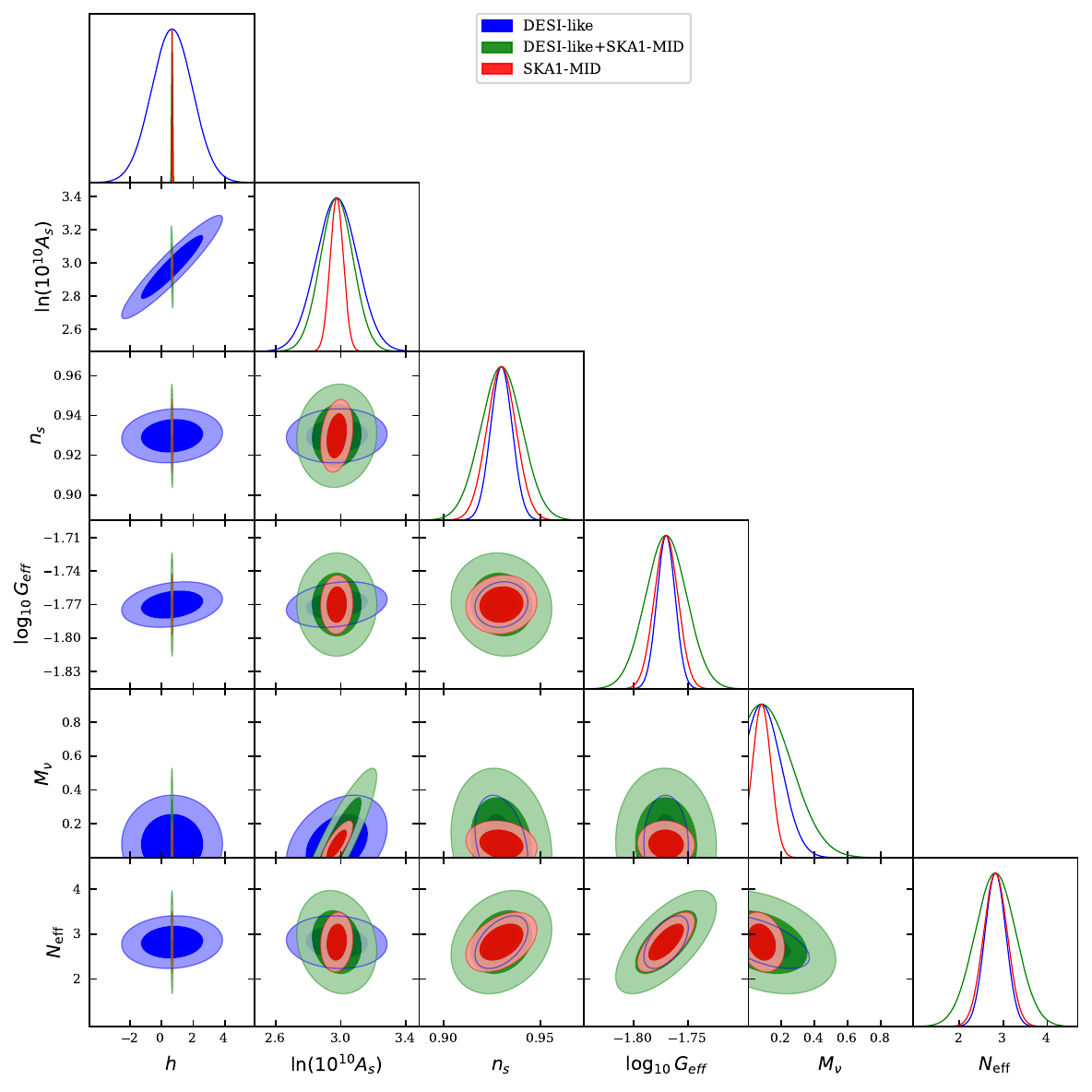}}
    \subfloat{\includegraphics[width=0.55\textwidth]{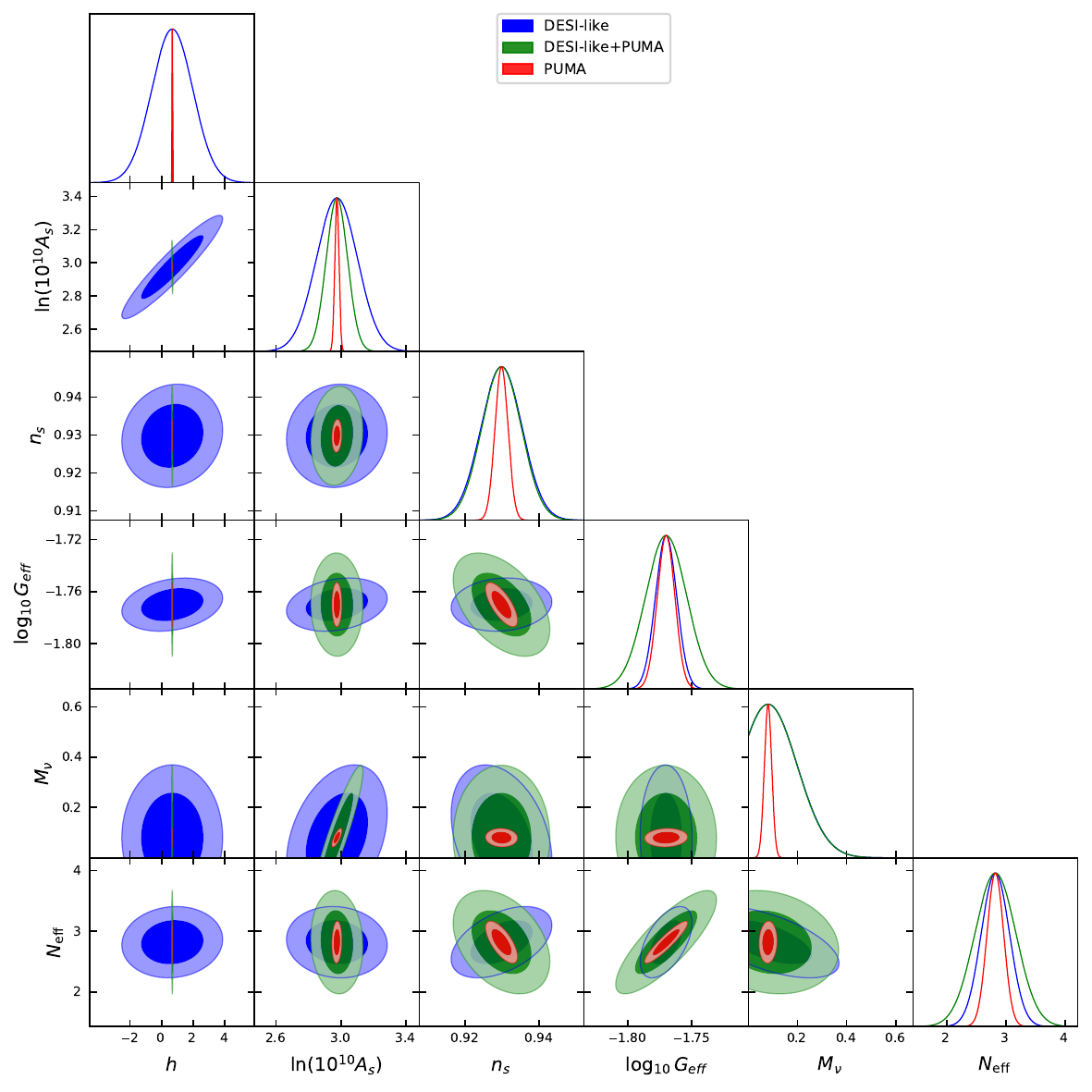}}
    \caption{Marginalized $1\sigma$ and $2\sigma$ Fisher contours in the SI$_\nu$ regime for the parameter set $\{h,\,\ln(10^{10}A_s),\,n_s,\,\log_{10}G_{\rm eff},\,M_\nu,\,N_{\rm eff}\}$ at the fiducial $\log_{10}(G_{\rm eff}/{\rm MeV}^{-2}) = -1.77$, from Ly$\alpha$ auto (blue), 21-cm auto (red) and Ly$\alpha$-21-cm cross spectrum (green) without CMB priors. The \textit{left panel} combines the DESI-like Ly$\alpha$ auto-spectrum with the SKA1-Mid 21-cm auto- and cross-spectra; the \textit{right panel} replaces SKA1-Mid with PUMA. The joint DESI-like+SKA1-Mid and DESI-like+PUMA analyses partially break the parameter degeneracies between $A_s$, $n_s$, and $G_{\mathrm{eff}}$, the marginalized error bars remain quite large. Notably, the Ly$\alpha$ auto-spectrum is quite insensitive to the Hubble parameter $h$, requiring cross-correlation with the 21-cm signal to constrain the expansion rate effectively.}
    \label{fig:SI_LSS_contours}
\end{figure}

\begin{figure}[ht!]
    \centering
    \subfloat{\includegraphics[width=0.55\textwidth]{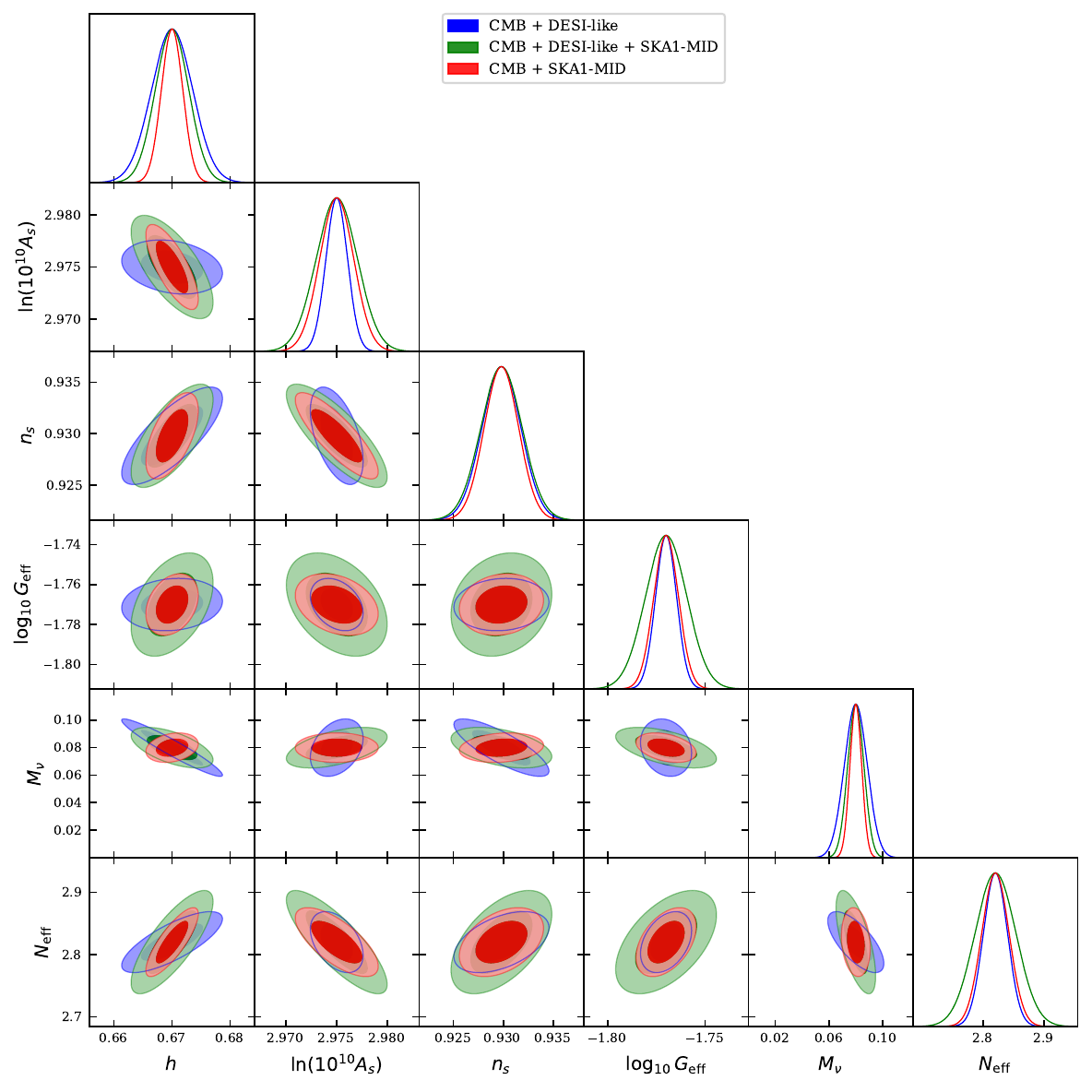}}
    \subfloat{\includegraphics[width=0.55\textwidth]{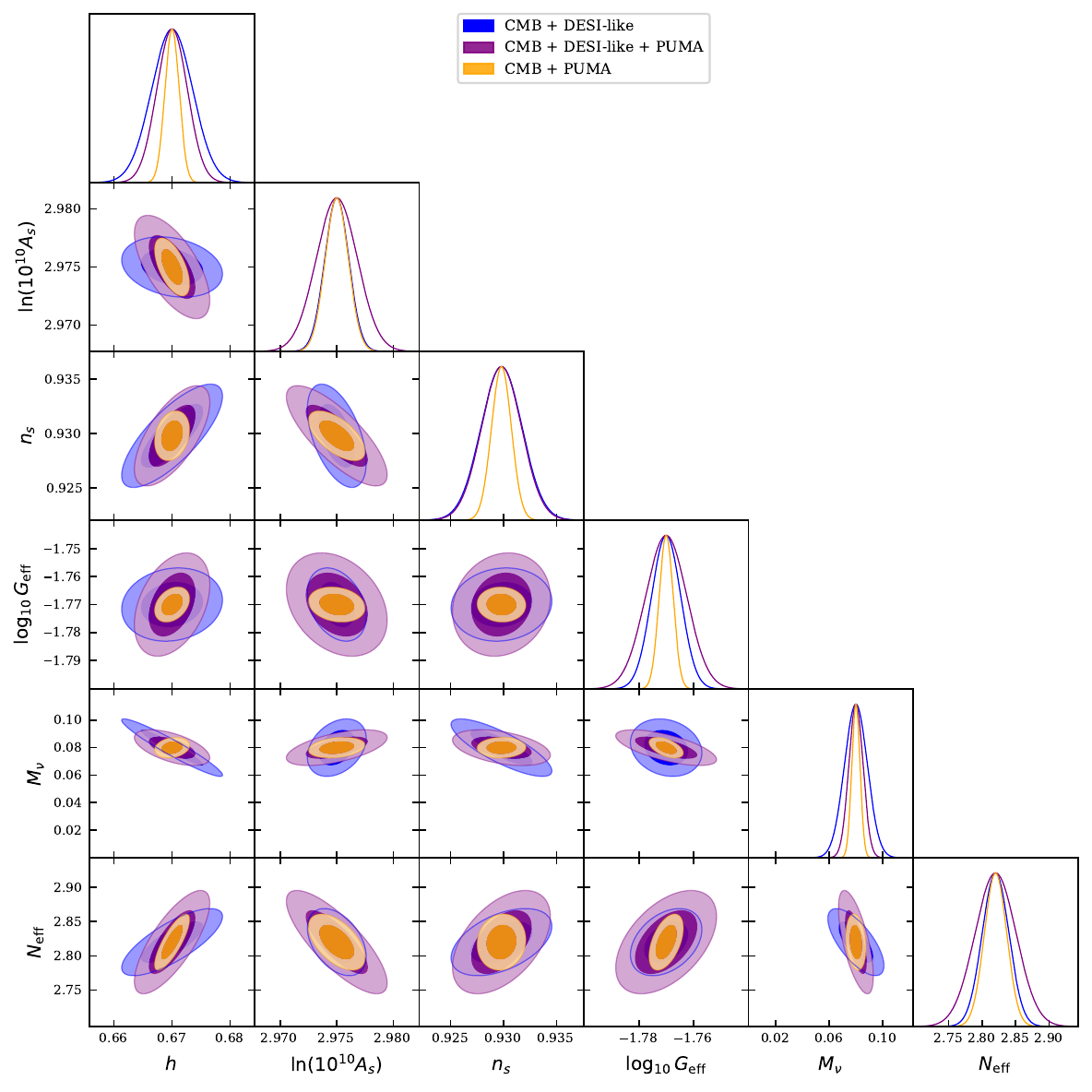}}
    \caption{Marginalized 1$\sigma$ and 2$\sigma$ Fisher contours in the SI$_\nu$ regime (at the fiducial $\log_{10}(G_{\rm eff}/{\rm MeV}^{-2}) = -1.77$) after incorporating CMB baseline priors. Contours represent constraints from the DESI-like Ly$\alpha$ auto-spectrum (blue), the 21-cm auto-spectrum (red), and the Ly$\alpha$--21-cm cross-spectrum (green), evaluated using SKA1-Mid (\textit{left}) and PUMA (\textit{right}) instrumental specifications in combination with CMB. CMB+PUMA achieves $\sigma(\log_{10}G_{\rm eff}) = 0.0026$, compared to $0.0063$ for CMB+SKA1-Mid, while the neutrino mass constraint tightens from $\sigma(M_\nu) = 0.0044\,\rm eV$ to $0.0030\,\rm eV$ over the same comparison.}
    \label{fig:SI_CMB_contours}
\end{figure}

\begin{figure}[ht!]
    \centering
    \includegraphics[width=1.0\linewidth]{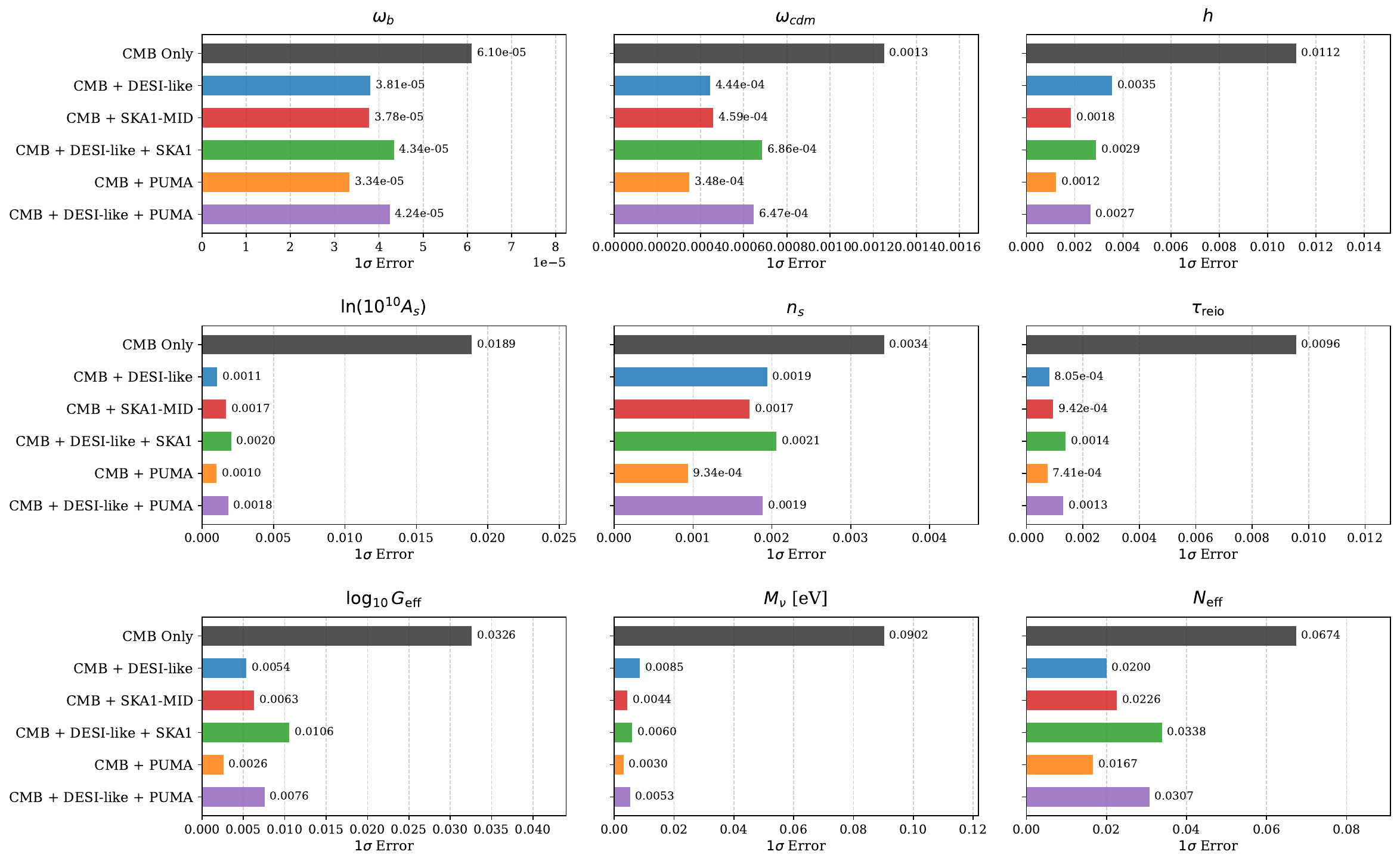}
    \caption{Marginalized $1\sigma$ forecast uncertainties on all nine parameters in the SI$_\nu$ regime across the six different combinations of observations with $\log_{10}G_
    {\rm eff}=-1.77$. The largest improvements over the CMB-only baseline are pronounced in $\ln(10^{10}A_s)$, $\log_{10}G_{\rm eff}$, and $M_\nu$, with CMB+PUMA providing the tightest individual constraints in each case. The joint DESI-like+PUMA combination does not meaningfully outperform PUMA alone for $\log_{10}G_{\rm eff}$ and $M_\nu$, confirming that the dominant constraining power in the SI$_\nu$ regime resides in the 21-cm rather than the Ly$\alpha$ auto-spectrum at the scales where the interaction signal is most pronounced.}
    \label{fig:SI_gantt}
\end{figure}

The Fisher posterior distributions and correlation ellipses for the parameters of interest $\{h, \ln(10^{10} A_s), n_s, \log_{10} G_{\mathrm{eff}}, M_\nu, N_{\mathrm{eff}}\}$ are shown in Fig.~\ref{fig:SI_LSS_contours} for all three observables based on the DESI-like combined with SKA1-Mid or PUMA combinations. For our analysis, we have taken 30 logarithmically equispaced $k$-bins in $k \in [0.01, 2] \, h \,{\mathrm{Mpc}}^{-1}$, and 10 equispaced $\mu$-bins in $\mu \in [0, 1]$. For the redshift distribution, we utilize 10 equispaced $z$-bins in $z \in [2.0, 3.05]$ for SKA1-Mid, and 15 equispaced $z$-bins in $z \in [2.0, 3.5]$ for PUMA. It should be noted that the Fisher contours featuring these specific parameters of interest are shown here, while the full parameter Fisher matrices are presented in the Appendix~\ref{app:9params}.

Let us now turn to the discussion of a few important features that immediately become apparent from Fig.~\ref{fig:SI_LSS_contours}. Looking first at the posteriors derived strictly from the Ly$\alpha$ and 21-cm observables (without CMB priors), the most striking feature is the shape of the contours in the $\ln(10^{10} A_s)$--$\log_{10} G_{\mathrm{eff}}$ plane. Because both parameters similarly dictate the amplitude of the late-time matter power spectrum, an enhanced interaction strength is easily offset by a reduced primordial amplitude. While introducing 21-cm data physically breaks this degeneracy, this allows the joint DESI-like+SKA1-Mid and DESI-like+PUMA analyses to disentangle $A_s$, $n_s$, and $G_{\mathrm{eff}}$. Nevertheless, without the inclusion of CMB priors, the marginalized error bars for these combined LSS tracers remain quite large.
Notably, the Ly$\alpha$ auto-spectrum is quite insensitive to the Hubble parameter $h$, requiring cross-correlation with the 21-cm signal to constrain the expansion rate effectively. However, PUMA already performs significantly better than SKA1-Mid in this regime because, its denser baseline coverage pushes sensitivity into the mildly non-linear scales where the SI$_\nu$ enhancement bump at $k \sim 0.2\text{--}0.5 \, h/\mathrm{Mpc}$ is most distinctive. Although it is important to note that the 21-cm auto spectrum and the cross spectrum  with SKA1-Mid may break the degeneracies between $n_s$ and $G_{\mathrm{eff}}$.
Furthermore, these combined probes demonstrate a strong capacity to constrain the sum of neutrino masses, $M_\nu$, with relatively high accuracy even before future CMB data is introduced.

The picture changes dramatically once the future CMB baseline  priors are included in the analysis, as shown in Fig.~\ref{fig:SI_CMB_contours}. Because the CMB independently pins down $A_s$ and $n_s$ through the temperature and polarization power spectra, the constraints on the parameters are tighter than the post-EoR tracers only analysis. The residual constraining power of the 21-cm and Ly$\alpha$ data is then directed almost entirely toward $G_{\mathrm{eff}}$, and the improvement relative to the CMB-only baseline is substantial. For the interaction coupling itself, CMB+PUMA reaches $\sigma(\log_{10} G_{\mathrm{eff}}) = 0.0026$, a factor of roughly 12 improvement over the CMB-only value of 0.033. The CMB+SKA1-Mid framework achieves $\sigma(\log_{10} G_{\mathrm{eff}}) = 0.0063$—better than the CMB alone by a factor of 5, but about 2.4 times weaker than CMB+PUMA. The difference comes down to scale coverage: SKA1-Mid's single-dish noise floor rises steeply at $k \gtrsim 0.3 \, h/\mathrm{Mpc}$, precisely where the SI$_\nu$ signal is largest, while PUMA's interferometric design maintains high sensitivity well into this regime.

A comprehensive overview of the projected uncertainties across the full parameter vector is visualized in the Gantt chart of Fig.~\ref{fig:SI_gantt}. The most dramatic statistical gains over the CMB-only baseline accumulate in $\ln(10^{10} A_s)$, which improves by a factor of 11 to 19 depending on the survey combination, ultimately reaching $\sigma(\ln(10^{10} A_s))=0.0010$ for the joint CMB+PUMA framework. Neutrino mass constraints also tighten drastically, dropping from $\sigma(M_\nu)=0.090$~eV with the CMB alone to $0.0030$~eV with CMB+PUMA and $0.0044$~eV with CMB+SKA1-Mid. These respective improvements by factors of $\sim 30$ and $\sim 20$ provide the statistical power necessary to cleanly distinguish the SI$_\nu$ fiducial mass of $0.08$~eV from the normal hierarchy minimum. Furthermore, CMB+PUMA achieves a tight constraint of $\sigma(N_{\rm eff})=0.0167$. This exceptional precision provides the statistical power necessary to distinguish the altered SI$_\nu$ fiducial value of 2.82 from the standard $\Lambda$CDM expectation of 3.044 at more than 13$\sigma$ significance.

One pattern that stands out in Fig.~\ref{fig:SI_gantt} is that the joint DESI-like+PUMA combination does not improve upon PUMA alone for $\log_{10}G_{\rm eff}$ and $M_\nu$. This implies that the Ly$\alpha$ auto-spectrum contributes mainly to constraining the broad-band shape parameters, $\ln(10^{10}A_s)$, and $n_s$, through its sensitivity to large-scale clustering, but its reach at the scales where the SI$_\nu$ interaction bump lives is blinded by aliasing noise and FoG damping at $k \gtrsim 0.3\,h \, {\rm Mpc}^{-1}$. On the other hand, the 21-cm auto- and cross-spectra carry the dominant signal at precisely those scales, so the Ly$\alpha$ forest acts more as a complementary anchor on standard cosmological parameters than as an independent probe of $G_{\rm eff}$ in this regime.

\subsubsection{MI$_\nu$ Mode}
\label{subsubsec:MI_results}
\begin{figure}[ht!]
    \centering
    \subfloat{\includegraphics[width=0.55\textwidth]{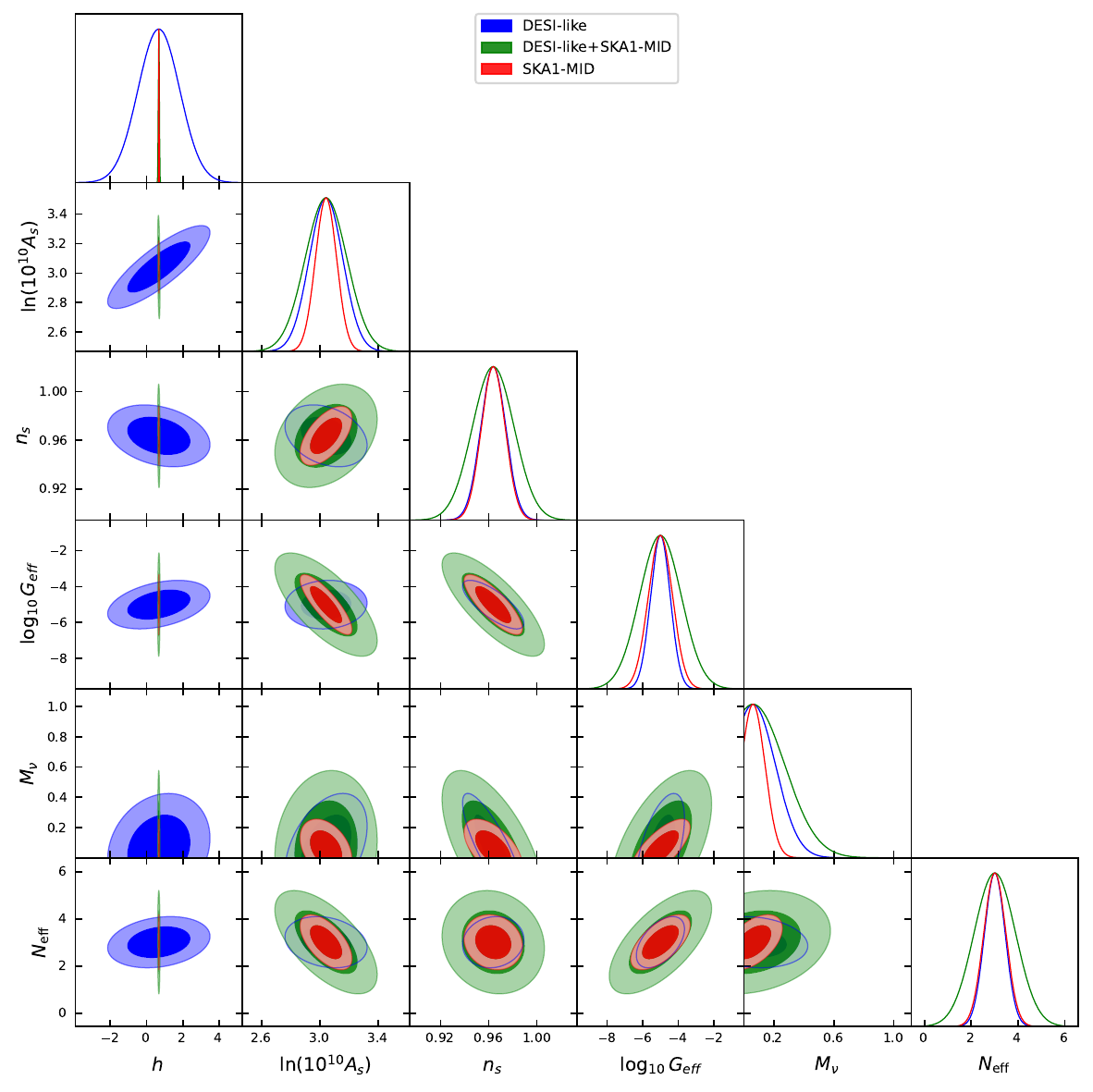}}
    \subfloat{\includegraphics[width=0.55\textwidth]{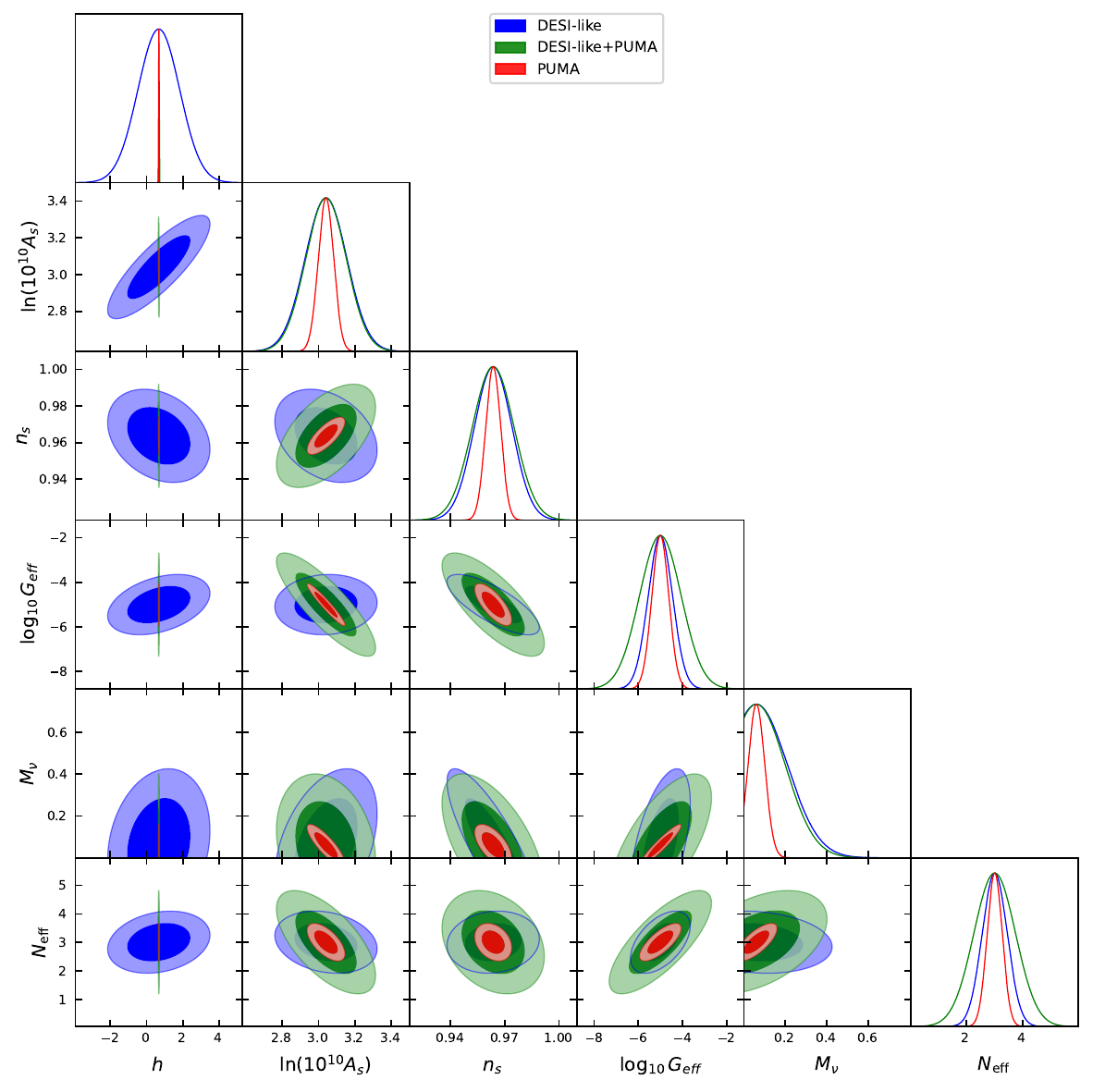}}
    \caption{Marginalized $1\sigma$--$2\sigma$ Fisher contours in the MI$_\nu$ regime for the parameter set $\{h,\,\ln(10^{10}A_s),\,n_s,\,\log_{10}G_{\rm eff},\,M_\nu,\,N_{\rm eff}\}$ at the fiducial $\log_{10}(G_{\rm eff}/{\rm MeV}^{-2}) = -5$, from Ly$\alpha$ auto (blue), 21-cm auto (red) and Ly$\alpha$-21-cm cross spectrum (green) without CMB priors. The \textit{left panel} shows constraints from DESI-like, SKA1-Mid, and their combination; the \textit{right panel} replaces SKA1-Mid with PUMA. The $\log_{10}G_{\rm eff}$ posteriors are vastly broader than in the SI$_\nu$ case, with the DESI-like contours spanning nearly the entire prior range along $\log_{10}G_{\rm eff}$, reflecting weaker signal at these weaker couplings.}
    \label{fig:MI_LSS_contours}
\end{figure}

\begin{figure}[ht!]
    \centering
    \subfloat{\includegraphics[width=0.55\textwidth]{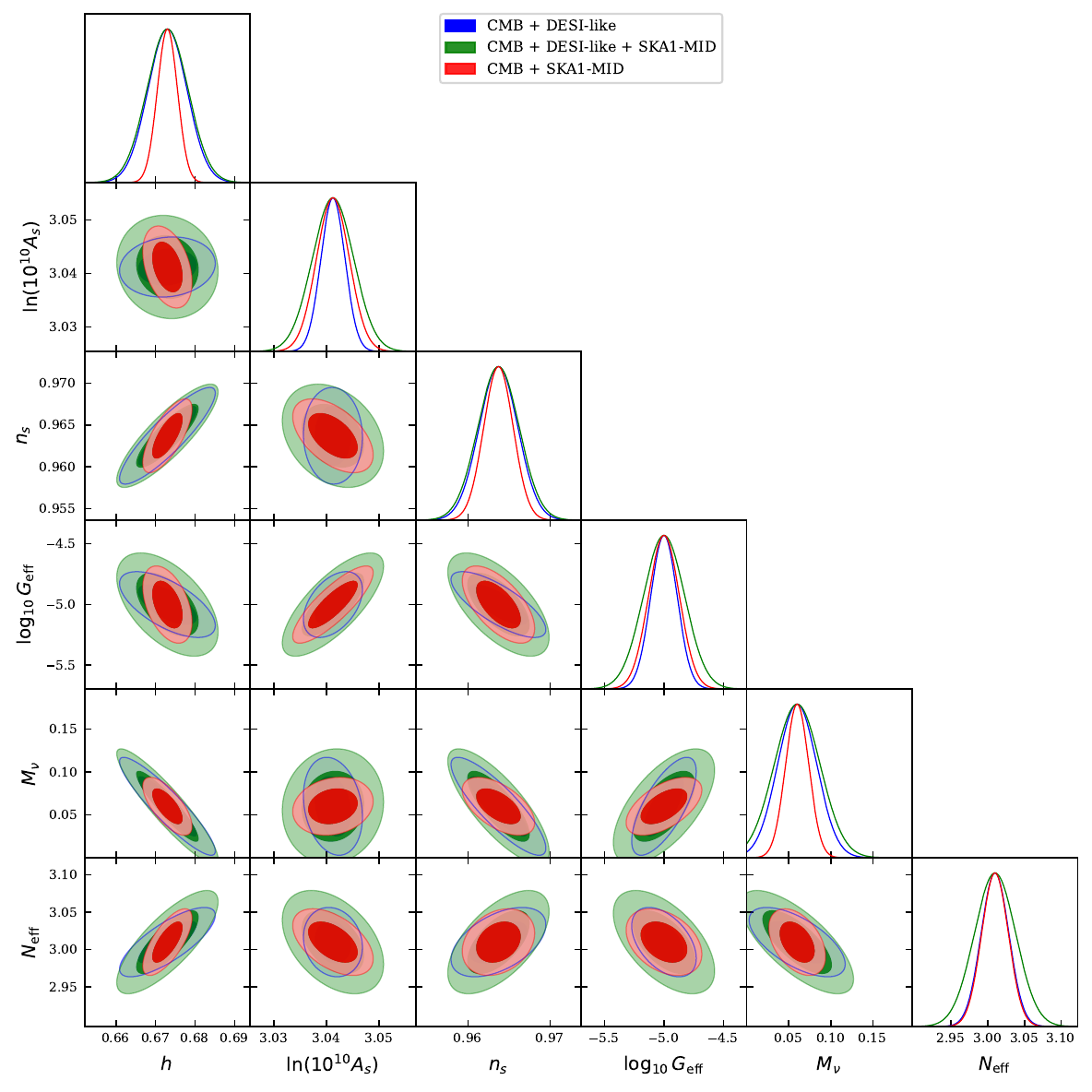}}
    \subfloat{\includegraphics[width=0.55\textwidth]{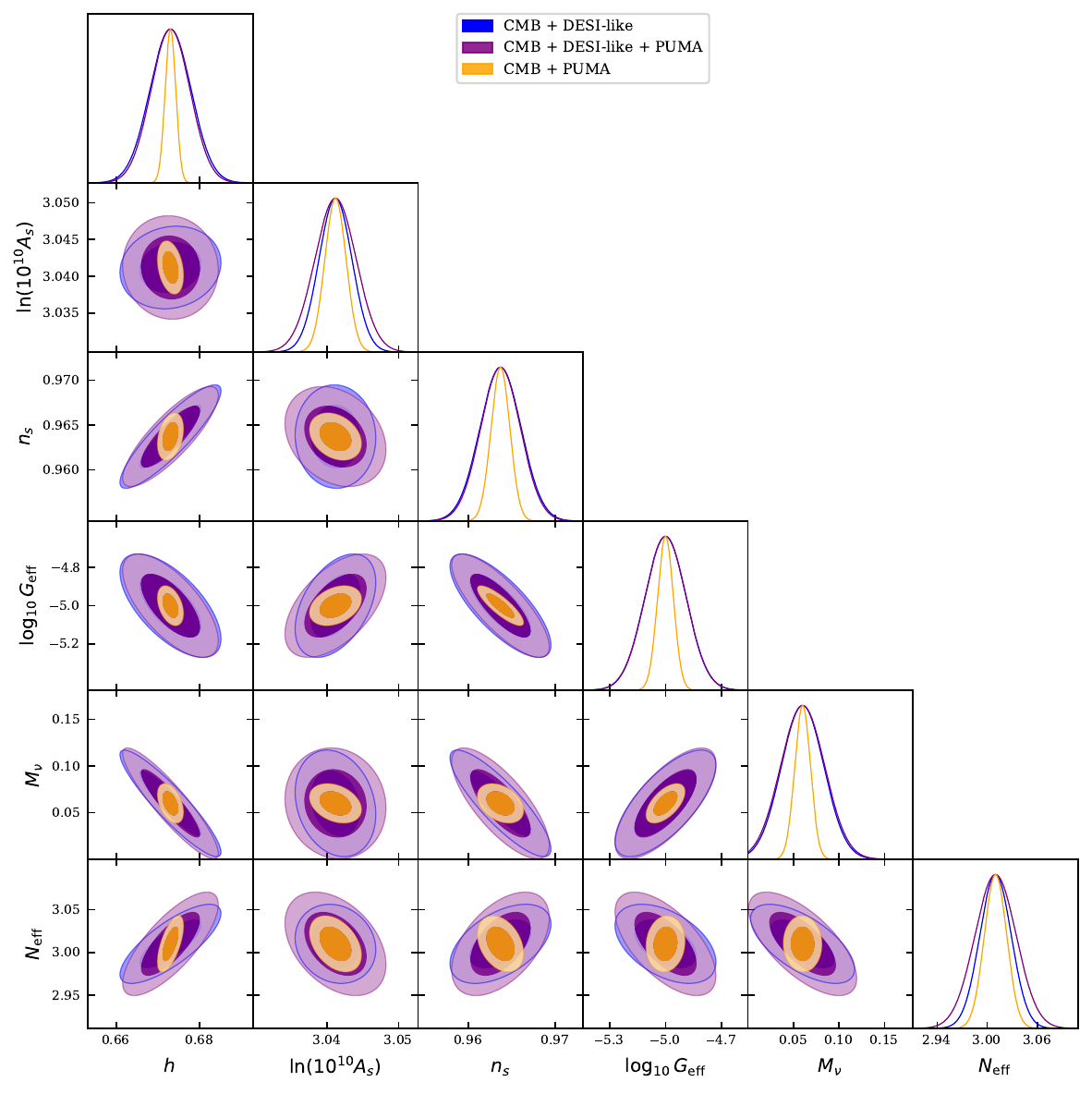}}
    \caption{Marginalized $1\sigma$--$2\sigma$ Fisher contours in the MI$_\nu$ regime after combining CMB baseline with the Ly$\alpha$ auto, 21-cm auto and  their cross-correlation, for SKA1-Mid (left) and PUMA (right). Unlike the SI$_\nu$ case, adding CMB does not improve all degeneracies equally: the $\log_{10}G_{\rm eff}$--$M_\nu$ plane retains residual correlation even after CMB priors are included, because the MI$_\nu$ signal lives at scales where both parameters independently suppress  small-scale power. The contours for CMB+PUMA are visibly tighter than CMB+SKA1-Mid across all parameters, with the most pronounced improvement along $\log_{10}G_{\rm eff}$ and $M_\nu$.}
    \label{fig:MI_CMB_contours}
\end{figure}

\begin{figure}[ht!]
    \centering
    \includegraphics[width=1.0\linewidth]{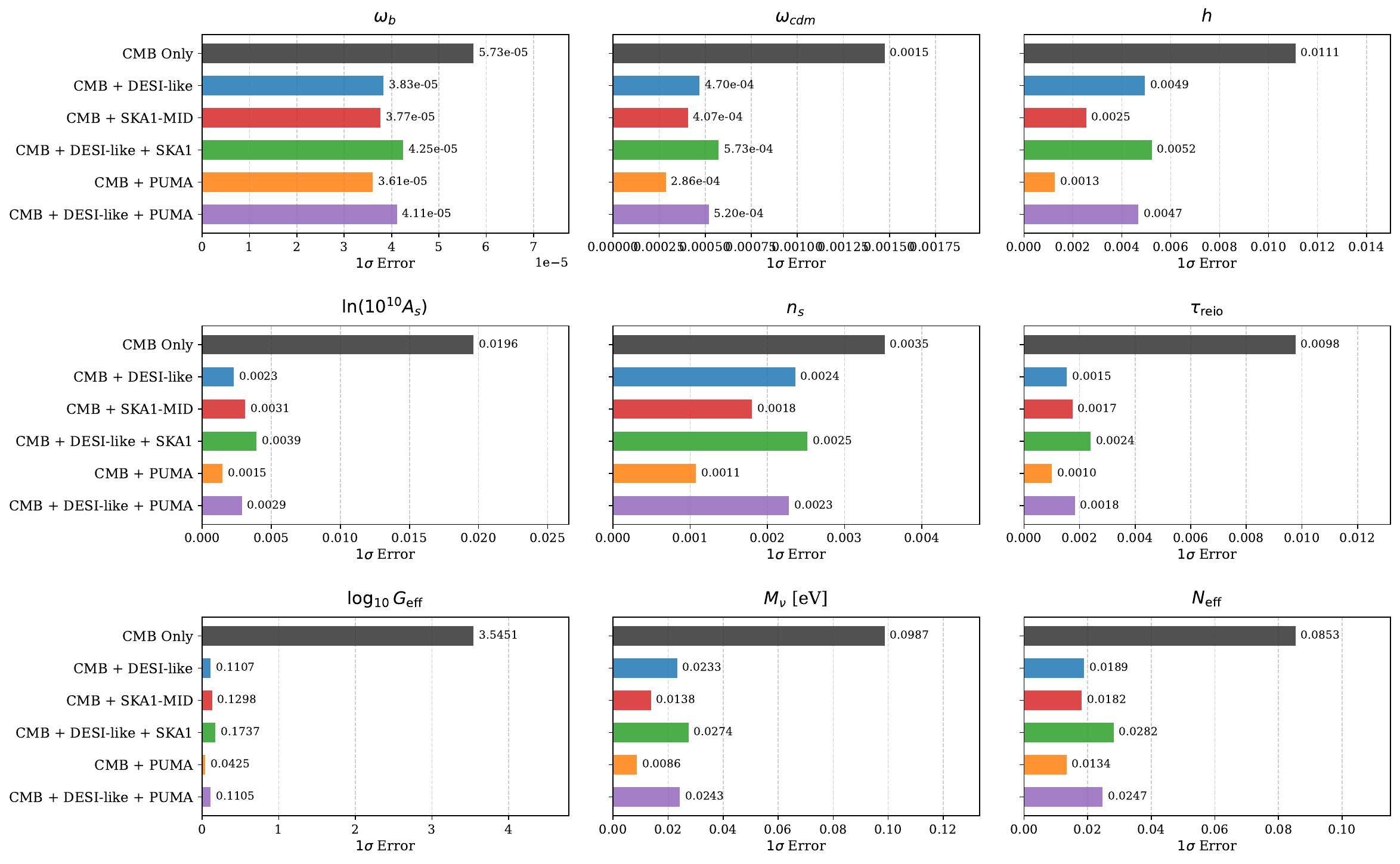}
    \caption{Marginalized $1\sigma$ forecast uncertainties on all nine parameters in the MI$_\nu$ regime across the six different combinations of observations with $\log_{10}G_{\rm eff}=-5$. The most striking feature is the CMB-only constraint on $\log_{10}G_{\rm eff}$, which is essentially uninformative at $\sigma(\log_{10}G_{\rm eff}) = 3.55$, since the MI$_\nu$ signal leaves no detectable imprint on CMB scales. The LSS surveys recover this situation dramatically: CMB+PUMA reaches $\sigma(\log_{10}G_{\rm eff}) = 0.043$, while CMB+SKA1-Mid achieves $0.130$. In contrast to the SI$_\nu$ case, the DESI-like survey  alone contributes meaningfully to $\log_{10}G_{\rm eff}$, and the  joint DESI-like+PUMA combination outperforms PUMA alone for $M_\nu$ and $N_{\rm eff}$, reflecting the more complementary roles of the two probes in this regime.}
    \label{fig:MI_gantt}
\end{figure}

The MI$_\nu$ mode presents a parameter forecasting landscape that contrasts sharply with the SI$_\nu$ regime. For the present analysis, the fiducial value of $G_{\rm eff}$ for the MI$_\nu$ mode is fixed to $\log_{10}G_{\rm eff}=-5$. Focusing initially on the large-scale structure (LSS) observables without CMB priors as shown in Fig.~\ref{fig:MI_LSS_contours}, the DESI-like posterior for the interaction coupling $\log_{10}G_{\rm eff}$ is practically unconstrained, flattening out. This behavior arises because the MI$_\nu$ signal manifests primarily as a high-$k$ enhancement (specifically at scales $k \gtrsim 1 \, h/\mathrm{Mpc}$), a region where the Ly$\alpha$ forest is heavily damped by aliasing noise and FoG suppression. Because the MI$_\nu$ fiducial model shares the same amplitude of $A_s$ and $n_s$ as the standard cosmological model, there is no corresponding large-scale deviation. Even the degeneracies of these parameters with $\log_{10}G_{\rm eff}$ are not significantly improved. 
However, Even operating independently of CMB data, PUMA yields a tightly bounded $\log_{10}G_{\rm eff}$ posterior that significantly outperforms both the SKA1-Mid and DESI-like configurations, owing to its interferometric design.

Once the future CMB baseline is added, as shown in Fig.~\ref{fig:MI_CMB_contours}, the overall picture tightens considerably, but the pattern of residual degeneracies differs from what we saw in the SI$_\nu$ case. There, CMB broke virtually all parameter correlations by pinning $A_s$. Here, the $\log_{10}G_{\rm eff}$--$M_\nu$ plane retains visible residual correlation even after CMB priors are included, because both parameters independently suppress small-scale matter clustering and the CMB does not directly probe the scales where either effect is dominant. The PUMA contours remain visibly superior to SKA1-Mid across all parameter pairs in Fig.~\ref{fig:MI_CMB_contours}, and the difference is most pronounced along the $\log_{10}G_{\rm eff}$ and $M_\nu$ directions, exactly where the high-$k$ sensitivity matters most. As in the SI$_\nu$ case, the sensitivity of Ly$\alpha$ auto spectrum to the Hubble parameter is insignificant, which improves once the 21-cm observables are added.

In Fig.~\ref{fig:MI_gantt}, we outline the errors on all the parameters through a Gantt chart for the MI$_\nu$ mode. The CMB-only error bar for $\log_{10}G_{\rm eff}$ stretches to $3.55$, effectively uninformative constraint, since the MI$_\nu$ signal produces no detectable imprint on CMB temperature or polarization at any multipole accessible to next-generation CMB surveys.  Adding DESI-like reduces this to $0.111$, a factor of $\sim 32$ improvement, but the most significant gain comes from PUMA. CMB+PUMA reaches $\sigma(\log_{10}G_{\rm eff}) = 0.043$, nearly a factor of three better than CMB+SKA1-Mid at $0.130$, and a factor of $\sim 83$ better than CMB alone. This is the single largest relative gain of any parameter combination across both interaction regimes studied in this work, and it underscores how uniquely suited PUMA is to the MI$_\nu$ problem.

The neutrino mass and $N_{\rm eff}$ constraints follow a similar but less extreme pattern. CMB+PUMA achieves $\sigma(M_\nu) = 0.0086\,\rm eV$, compared to $0.0138\,\rm eV$ for CMB+SKA1-Mid and $0.0987\,\rm eV$ for CMB alone -- roughly a factor of 11 improvement over CMB only. For $N_{\rm eff}$, the corresponding numbers are $0.013$ (CMB+PUMA), $0.018$ (CMB+SKA1-Mid), and $0.085$ (CMB only). One feature worth highlighting is that, unlike in the SI$_\nu$ case, the joint DESI-like+PUMA combination does offer a modest improvement over PUMA alone for $M_\nu$ and $N_{\rm eff}$, going from $0.0086$ to $0.0243\,\rm eV$ and from $0.013$ to $0.025$ respectively when DESI-like is included alongside PUMA. This reflects the more genuinely complementary role of the two surveys in the MI$_\nu$ regime. The Ly$\alpha$ forest contributes useful information about the broad-band shape of the matter power spectrum at intermediate scales, which helps break the residual $M_\nu$--$N_{\rm eff}$ degeneracy that PUMA alone cannot fully resolve. The interaction coupling itself, however, remains dominated by PUMA, with the joint combination offering no improvement over PUMA alone for $\log_{10}G_{\rm eff}$.

\begin{figure}[ht!]
    \centering
    \includegraphics[width=0.8\linewidth]{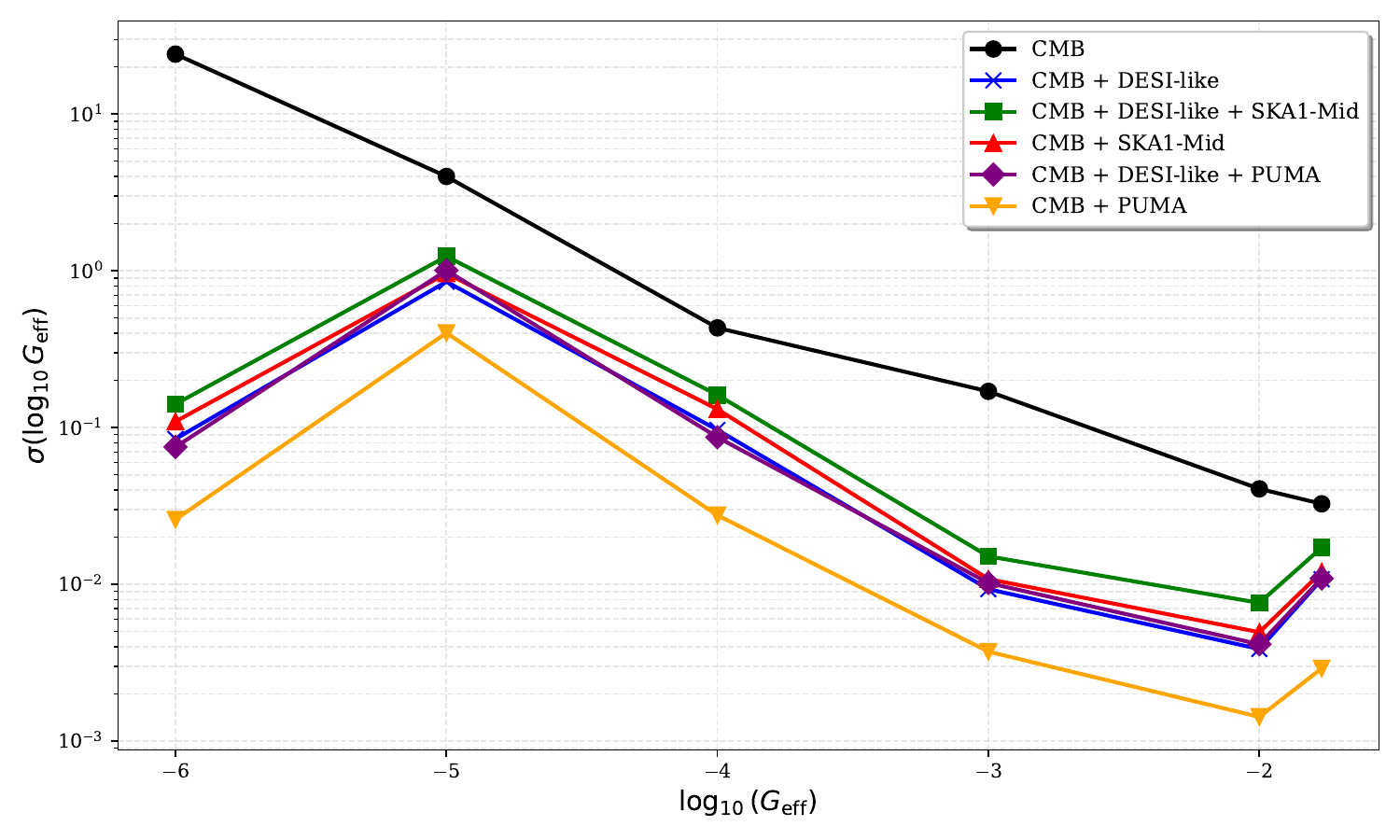}
    \caption{Marginalized $1\sigma$ forecast uncertainty on $\log_{10}G_{\rm eff}$ as a function of the fiducial coupling strength. The curves contrast the constraining power of six data combinations: CMB alone (black), CMB + DESI-like Ly$\alpha$ auto-spectrum (blue), CMB + SKA1-Mid cross-spectrum (green), CMB + SKA1-Mid 21-cm auto-spectrum (red), CMB + PUMA cross-spectrum (purple), and CMB + PUMA 21-cm auto-spectrum (orange). All configurations exhibit a broad sensitivity maximum (corresponding to minimum forecast uncertainty) in the SI$_\nu$ regime ($\log_{10}(G_{\rm eff}/{\rm MeV}^{-2}) \approx -3$ to $-1.77$), where the interaction-induced clustering enhancement aligns with the optimal scales of the corresponding survey. Conversely, a pronounced degradation in sensitivity occurs near $\log_{10}G_{\rm eff} \approx -5$ as the signal is mainly amplified near the highly non-linear scales. Notably, the PUMA 21-cm auto-spectrum (orange) provides the most stringent constraints uniformly across the parameter space, uniquely maintaining $\sigma(\log_{10}G_{\rm eff}) \lesssim 0.1$ even at the weakest explored couplings.}
    \label{fig:Geff_sensitivity}
\end{figure}

The results discussed above were each evaluated at fixed fiducial values of $G_{\rm eff}$, and it is worth asking how representative those choices are across the broader parameter space. Fig.~\ref{fig:Geff_sensitivity} represents $\sigma(\log_{10}G_{\rm eff})$ for different values of the fiducial coupling strength from $\log_{10}(G_{\rm eff}/{\rm MeV}^{-2}) = -6$ all the way to $-1.77$, allowing a direct comparison of all six observations across the full MI$_\nu$--SI$_\nu$ range.

In the weak-coupling regime ($\log_{10}G_{\rm eff} = -6$), the forecasted uncertainties across all tracer observables span roughly an order of magnitude. The PUMA 21-cm auto-spectrum provides the most stringent constraints, achieving $\sigma(\log_{10}G_{\rm eff}) \approx 0.025$, whereas combinations involving SKA1-Mid and DESI-like surveys yield uncertainties in the range of $\sigma \approx 0.08 -0.15$. As the coupling strength approaches the moderately interacting (MI$\nu$) fiducial at $\log_{10}G_{\rm eff} = -4$, the constraining power of the PUMA 21-cm observable remains dominant ($\sigma \approx 0.025$), while the cross-spectrum and DESI-like combinations asymptote to $\sigma \approx 0.08-0.13$. 
The PUMA 21-cm auto-spectrum achieves an optimal sensitivity of $\sigma \approx 0.001$, and even the least sensitive combinations reach $\sigma \approx 0.004$--$0.005$ in the strongly interacting (SI$\nu$) regime ($\log_{10}G_{\mathrm{eff}} \in [-2, -1.77]$), consistent with the fixed-fiducial results presented in Section~\ref{subsubsec:SI_results}.

Two more specific features are worth noting. Across the entire coupling range, the PUMA 21-cm auto-spectrum outperforms the PUMA cross-spectrum by a factor of roughly 3--10. Within a Fisher framework this reflects the signal-to-noise advantage of the 21-cm auto at the relevant scales, though in practice the cross-spectrum's immunity to single-tracer systematics and foreground contamination makes it the more robust observable for real data analyses. Separately, the SKA1-Mid cross and SKA1-Mid 21-cm auto error bars closely follow each other  across the full coupling range, as do DESI-like and SKA1-Mid cross at weak couplings, suggesting that at moderate interaction strengths the constraining power is roughly equally across the Ly$\alpha$ and 21-cm.
Overall, the figure confirms that the qualitative conclusions drawn from the two fixed-fiducial analyses, PUMA's dominance across all regimes and the broad accessibility of the SI$_\nu$ regime, hold uniformly across the full range of coupling strengths explored in this work.

\subsection{Impact of Foregrounds on 21-cm Observables}
\label{sec:foregrounds}
While the 21-cm auto-spectrum holds immense statistical power in an idealized scenario, detecting the cosmological signal requires subtracting astrophysical foregrounds—such as Galactic synchrotron emission and extragalactic point sources—that are several orders of magnitude brighter~\cite{Cunnington:2020wdu,Cunnington:2021czb,Spinelli:2021emp,Cunnington:2023jpq}. Furthermore, the finite angular resolution of the observing instrument inherently filters the detectable signal. To ensure our analysis yields realistic constraints, we introduce a phenomenological observational quantity, $\mathcal{E}_{21}$, which models both the mode loss inherent to foreground removal and the specific telescope beam effects following~\cite{Kopana:2024qqq}:
\begin{equation}
\mathcal{E}_{21}(k_{\parallel}, k_{\perp}, z) = \mathcal{S}_{\mathrm{fg}}(k_{\parallel}, k_{\perp}, z) \times \mathcal{B}_{\mathrm{mode}}(k_{\perp}, z)\,.
\end{equation}

The first component, $\mathcal{S}_{\mathrm{fg}}$, captures the signal attenuation due to foreground cleaning. It can be decomposed into two distinct geometric effects in Fourier space:
\begin{equation}
\mathcal{S}_{\mathrm{fg}}(k_{\parallel}, k_{\perp}, z) = \mathcal{F}_{\mathrm{rad}}(k_{\parallel}) \times \mathcal{F}_{\mathrm{wedge}}(k_{\parallel}, k_{\perp}, z)\,.
\end{equation}

Here, $\mathcal{F}_{\mathrm{rad}}$ accounts for radial foreground damping. Because foreground emissions are spectrally smooth, removing them inevitably destroys cosmological information contained in long-wavelength line-of-sight modes (low $k_{\parallel}$). This effect impacts both single-dish and interferometric observations and is modeled as a Gaussian high-pass filter:
\begin{equation}
\mathcal{F}_{\mathrm{rad}}(k_{\parallel}) = 1 - \exp\left[ -\left( \frac{k_{\parallel}}{k_{\mathrm{fg}}} \right)^2 \right]\,,
\end{equation}
where $k_{\mathrm{fg}}$ represents the characteristic inverse scale of foreground contamination. For our analysis, we adopt a conservative cutoff of $k_{\mathrm{fg}} = 0.01 \ h \, \mathrm{Mpc}^{-1}$.

The term $\mathcal{F}_{\mathrm{wedge}}$ accounts for the mode-mixing introduced by the chromatic nature of an interferometer's primary beam, which scatters smooth foregrounds into high-$k_{\parallel}$ modes, creating a highly contaminated ``wedge.'' For interferometric arrays like PUMA, we apply a strict foreground avoidance strategy using a Heaviside step function:
\begin{equation}
\mathcal{F}_{\mathrm{wedge}}(k_{\parallel}, k_{\perp}, z) = \Theta\left( k_{\parallel} - W(z) k_{\perp} \right)\,.
\end{equation}
The boundary of this wedge is dictated by the geometry of the primary beam~\cite{Pober:2013jna,Pober:2014lva,Obuljen:2017jiy,Alonso:2017dgh,Ghosh:2017woo,Karagiannis:2019jjx,PUMA:2019jwd,Kopana:2024qqq}:
\begin{equation}
W(z) = r(z) H(z) \sin\left[ 0.61 N_w \theta_{\mathrm{FOV}}(z) \right]\, ,
\end{equation}
where $r(z)$ is the comoving radial distance, $H(z)$ is the conformal Hubble parameter, $\theta_{\mathrm{FOV}}(z)$ is the primary beam's field of view, and $N_w$ parameterizes the extent of the beam sidelobes leaking into the measurement. We assume $N_w = 1$. For single-dish configurations like SKA1-Mid, which do not suffer from this specific interferometric wedge effect, we set $\mathcal{F}_{\mathrm{wedge}} = 1$.

The second component of the observational envelope, $\mathcal{B}_{\mathrm{mode}}$, dictates the instrument's sensitivity to transverse scales ($k_{\perp}$). This factor is fundamentally different depending on the survey's operational mode. For an interferometer, the beam factor simply scales with the field of view:
\begin{equation}
\mathcal{B}_{\mathrm{int}} = \theta_{\mathrm{FOV}}\,.
\end{equation}

Conversely, a single-dish array suffers from an exponential dampening of transverse modes due to its finite physical aperture, resulting in a loss of small-scale angular resolution. We model this as a Gaussian beam profile:
\begin{equation}
\mathcal{B}_{\mathrm{sd}}(k_{\perp}, z) = \exp\left( - \frac{k_{\perp}^2 r(z)^2 \theta_{\mathrm{FOV}}^2}{16 \ln 2} \right)\,.
\end{equation}

To incorporate these realistic operational and astrophysical limitations into our signal-to-noise calculations, we modulate the theoretical power spectra. The observed 21-cm auto-spectrum is suppressed by the square of this envelope, $P_{21}^{\mathrm{obs}} = \mathcal{E}_{21}^2 P_{21}^{\mathrm{3D}}$, while the cross-power spectrum, correlating the 21-cm field with the foreground-independent Ly$\alpha$ flux, scales linearly with the envelope, $P_{21, F}^{\mathrm{obs}} = \mathcal{E}_{21} P_{21, F}$.
\begin{figure}[t]
\centering
\includegraphics[width=1.0\textwidth]{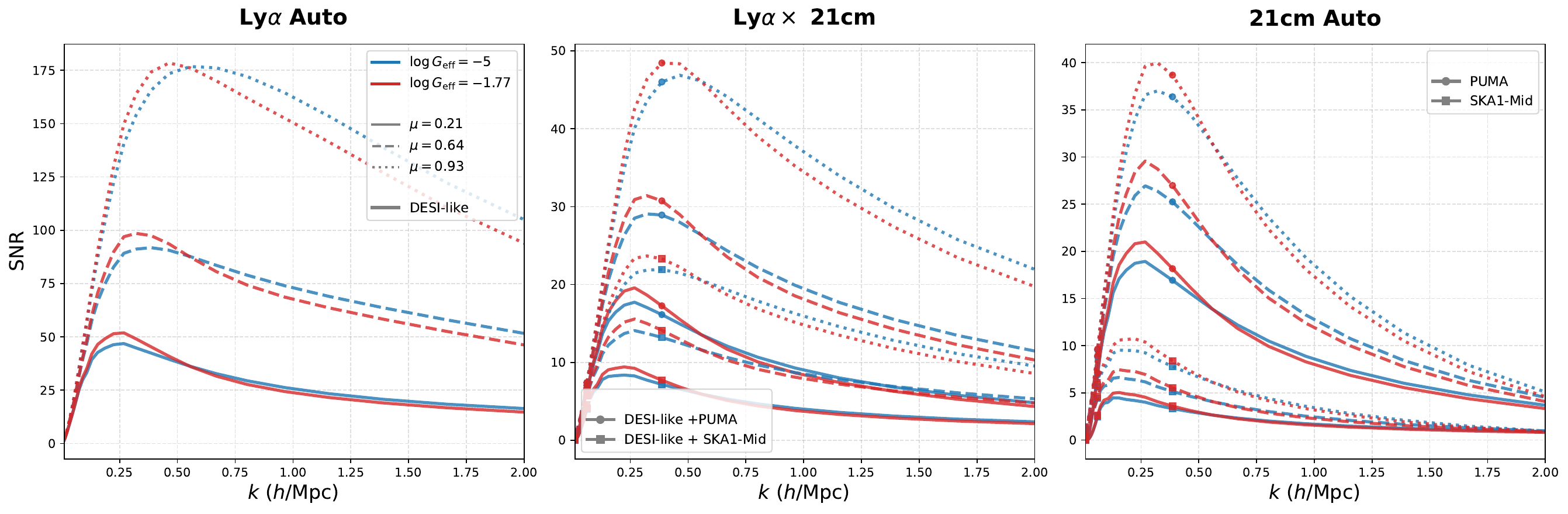}
\caption{Signal-to-noise ratio (SNR) as a function of scale, incorporating the realistic 21-cm foreground attenuation envelope ($\mathcal{E}_{21}$) with a radial damping cutoff of $k_{\mathrm{fg}} = 0.01 \ h\,\mathrm{Mpc}^{-1}$. The panels display the Ly$\alpha$ auto (\textit{left}), Ly$\alpha - 21$-cm cross (\textit{middle}), and 21-cm auto (\textit{right}) spectra for the MI$_\nu$ ($\log_{10} G_{\mathrm{eff}} = -5$, blue) and SI$_\nu$ ($\log_{10} G_{\mathrm{eff}} = -1.77$, red) modes at a fixed redshift $z = 2.33$. While the optical Ly$\alpha$ auto-spectrum remains unaffected by radio systematics, the inclusion of the foreground wedge and radial damping drastically suppresses the 21-cm auto-spectrum, dropping its peak SNR by nearly an order of magnitude and reducing it to levels highly comparable with the cross-spectrum.}
\label{fig:snr_vs_k_fg}
\end{figure}

To explicitly demonstrate the impact of these realistic observational constraints on our analysis, we compare the ideal, foreground-free signal-to-noise ratio with the scenario incorporating the signal attenuation envelope, $\mathcal{E}_{21}$.

In the idealized, foreground-free case, the 21-cm auto-spectrum overwhelmingly dominates the statistical sensitivity. Driven by the sheer volume and low thermal noise of the radio arrays, it reaches an optimistic SNR of $\approx 400$ for PUMA at optimal line-of-sight angles ($\mu=0.93$). Under these same conditions, the Ly$\alpha$--21-cm cross-spectrum and the Ly$\alpha$ auto-spectrum achieve maximum SNRs of roughly $175$.

However, the inclusion of realistic radio foregrounds completely upends this statistical hierarchy as depicted in Fig.~\ref{fig:snr_vs_k_fg}. Implementing the radial damping and the primary beam wedge severely penalizes the 21-cm observables. Because it relies on optical spectroscopy, the Ly$\alpha$ auto-spectrum remains entirely unaffected by radio foregrounds, retaining its peak SNR. In contrast, the 21-cm auto-spectrum undergoes a suppression at large and intermediate scales; its peak SNR plummets by nearly an order of magnitude, dropping to $\approx 40\text{--}50$.

Although the SNR of the Ly$\alpha$--21-cm cross-spectrum also decreases (to roughly 35–40), this drop is far less severe than the one suffered by the 21-cm auto-spectrum. This resilience occurs because radio foregrounds do not fundamentally bias the cross-correlation signal. Mathematically, the cross-spectrum only suffers a linear penalty from the signal attenuation compared to the quadratic attenuation factor in the 21-cm auto spectrum.
The Ly$\alpha$--21-cm cross-spectrum also experiences a reduction in SNR, falling to $\approx 35\text{--}40$. This occurs because, even though the cross-power signal itself is unbiased by purely radio foregrounds, the variance of the estimator is increased by the 21-cm foreground power present in the corresponding modes.

\section{Summary and Conclusion}
\label{sec:summary}

In this work we have investigated the signatures of early-universe neutrino self-interactions on post-reionization large-scale structure tracers and forecast the precision with which upcoming surveys can constrain the effective coupling $G_{\rm eff}$. Our analysis spans two qualitatively distinct interaction regimes, the strongly interacting SI$_\nu$ mode near $\log_{10}(G_{\rm eff}/{\rm MeV}^{-2}) = -1.77$ and the moderately interacting MI$_\nu$ mode near $\log_{10}(G_{\rm eff}/{\rm MeV}^{-2})=-5$. We present a  Fisher matrix forecast analysis for the Ly$\alpha$ auto spectrum, 21-cm auto spectrum, and their cross-correlation, considering future surveys such as a DESI-like instrument, SKA1-Mid, and PUMA, followed by a joint analysis incorporating priors from a representative future CMB mission like CMB-S4.
The theoretical framework rests on the effective four-fermion parameterization of neutrino self-interactions, implemented through a modified Boltzmann hierarchy solved with a custom version of CLASS, and the observable power spectra are modeled as biased tracers of the underlying matter field in the mildly non-linear regime including Kaiser redshift-space distortions and FoG damping.

The primary physical mechanism driving our results is the scale-dependent imprint of delayed neutrino free-streaming on the matter power spectrum. In the SI$_\nu$ regime, self-decoupling near matter-radiation equality produces a characteristic 10--15\% enhancement at $k \sim 0.2$--$0.5\,h \, {\rm Mpc}^{-1}$, accompanied by a 30--40\% suppression at smaller scales driven by the reduced $A_s$ and $n_s$ values the mode requires. Conversely, in the MI$_\nu$ regime, the enhancement bump shifts to $k \gtrsim 1\,h \, {\rm Mpc}^{-1}$ without any accompanying shift in the primordial parameters, making the signal entirely invisible to the CMB but directly accessible to high-resolution radio interferometry. The Ly$\alpha$--21-cm cross-correlation is particularly well suited to probe both regimes. Because the two signals are measured by independent instruments through entirely different physical processes, their instrumental systematics and foreground contaminations are statistically uncorrelated. However, realistic foreground estimation reduces the SNR to some extent, which is more pronounced in the case of the auto-spectra than the cross-spectrum, still leaving enough room for detection.

Our forecast analysis outline a consistent and physically transparent scenario. For the ${\rm SI}_\nu$ mode, Ly$\alpha$ data alone exhibit a slight degeneracy between $\ln(10^{10}A_s)$ and $\log_{10}G_{\rm eff}$, since both parameters pull on the overall normalization of the matter power spectrum in nearly identical ways. However, combining this with 21-cm intensity mapping data decisively breaks this degeneracy. The 21-cm data independently pin down $A_s$ through the linear-scale regime of the power spectra, after which the residual LSS constraining power is directed almost entirely toward $G_{\rm eff}$. It is important to note that in a CMB-only analysis, this parameter degeneracy persists. However, when combining these multi-tracer LSS constraints with future CMB data, the improvement in overall parameter precision is substantial.
The resulting constraints are striking: CMB+PUMA reaches $\sigma(\log_{10}G_{\rm eff}) = 0.0026$, a factor of $\sim 12$ improvement over CMB alone, and $\sigma(M_\nu) = 0.003\,\rm eV$, sufficient to distinguish the SI$_\nu$ fiducial neutrino mass from the normal hierarchy minimum with high significance. CMB+SKA1-Mid achieves comparable but weaker constraints, at $\sigma(\log_{10}G_{\rm eff}) = 0.0063$ and $\sigma(M_\nu) = 0.0044\,\rm eV$, reflecting SKA1-Mid's steeper single-dish noise floor at the scales where the SI$_\nu$ signal is most pronounced.

The MI$_\nu$ mode presents a more challenging but ultimately more rigorous picture. The CMB-only constraint on $\log_{10}G_{\rm eff}$ is effectively uninformative at $\sigma = 3.55$, since the MI$_\nu$ signal leaves no detectable imprint at any CMB multipole accessible to next-generation CMB surveys. Adding a DESI-like Ly$\alpha$ survey recovers a factor of $\sim 32$ improvement to $\sigma = 0.111$, but the MI$_\nu$ enhancement at $k \gtrsim 1\,h \, {\rm Mpc}^{-1}$ lies beyond the reach of optical spectroscopy alone, which is severely damped by aliasing noise and Finger-of-God suppression at these scales. PUMA resolves this directly: CMB+PUMA achieves $\sigma(\log_{10}G_{\rm eff}) = 0.043$, nearly two orders of magnitude improvement over CMB alone and nearly a factor of three better than CMB+SKA1-Mid. This represents the single largest relative gain of any parameter combination across both interaction regimes studied here, and underscores that PUMA is the only realistic path to a meaningful MI$_\nu$ constraint with near-future instrumentation. Unlike in the SI$_\nu$ case, the $\log_{10}G_{\rm eff}$--$M_\nu$ degeneracy persists in a more pronounced way, even after CMB priors are applied, because none of the parameter affects CMB scales directly in this regime, and breaking it requires the joint DESI-like+PUMA combination which improves the neutrino mass constraint to $\sigma(M_\nu) = 0.0086\,\rm eV$.

Furthermore, we evaluated the evolution of these constraints across the full coupling parameter space, spanning $\log_{10}G_{\rm eff} = -6$ to $-1.77$, demonstrating that our qualitative conclusions are robust and independent of the selected fiducial values. Every survey combination exhibits an optimal sensitivity (corresponding to minimum forecast uncertainty) within the SI$_\nu$ regime, followed by a pronounced degradation in constraining power near $\log_{10}G_{\rm eff} \approx -5$. This degradation arises because the interaction-induced clustering enhancement shifts to the boundary between the mildly and highly non-linear regimes, an intermediate scale where all considered instruments experience a simultaneous loss of sensitivity. Nevertheless, the PUMA 21-cm auto-spectrum consistently yields the most stringent constraints across the entire coupling range, representing the sole configuration capable of maintaining $\sigma(\log_{10}G_{\rm eff}) \lesssim 0.1$ even at the weakest interaction strengths explored.

Several directions for future work naturally follow from this analysis. Our forecasts rely on the Fisher matrix formalism and therefore represent optimistic lower bounds on the achievable uncertainties; a full MCMC analysis incorporating realistic foreground models, beam systematics, 
and non-Gaussian covariances would improve the picture considerably. The HI bias model adopted here is calibrated against hydrodynamical simulations at $z \leq 6$ and may underestimate contributions from low-mass halos at higher redshifts, introducing a potential source of systematic uncertainty that will need to be addressed as survey sensitivity improves. Finally, we have treated the two interaction regimes as independent analyses at fixed fiducials; a joint multi-fiducial forecast covering the full bimodal posterior in $G_{\rm eff}$ would provide a more complete assessment of the discriminating power of these probes between the SI$_\nu$ and MI$_\nu$ modes. These are left for future work.
 
In summary, the combination of post-reionization 21-cm intensity mapping and the Ly$\alpha$ forest, anchored by future CMB priors, provides a powerful and complementary window into BSM neutrino physics that is qualitatively different from what either probe can achieve alone. For the SI$_\nu$ mode, the cross-correlation breaks the degeneracies that have limited CMB-based analyses and delivers precision constraints on both $G_{\rm eff}$ and $M_\nu$. For the MI$_\nu$ mode, PUMA offers the first realistic prospect of a detection. As SKA1-Mid, PUMA move towards their design sensitivities over the coming decade, the observational program outlined here will be directly testable, and the combination of these datasets with representative future CMB mission like CMB-S4 has the potential to either firmly establish or rule out neutrino self-interactions in the early-universe.

\section*{Acknowledgments}

Authors acknowledge useful discussions with Arko Bhaumik, Paulo Montero-Camacho and Debarun Paul. The authors also thank the anonymous referee for valuable comments and insights that helped improve the manuscript.
SP1 thanks CSIR for financial support through Senior Research Fellowship (File no. 09/093(0195)/2020-EMR-I).
SP2 thanks ANRF,  Govt. of India, for partial support through Project No. CRG/2023/003984. We acknowledge the computational facilities provided by the SyMeC HPC cluster of ISI Kolkata.

\appendix

\section{CMB Noise Modeling and Covariance Matrix}
\label{app:cmb_noise}
As noted in Sec.~\ref{sec:introduction}, our study extends beyond upcoming post-EoR observables to provide forecasts for future CMB missions. Here, we outline the noise modeling incorporated into our analysis for upcoming CMB experiments, adopting CMB-S4 specifications as a representative baseline.
The noise properties of a CMB experiment are set by the map-level sensitivity and the angular resolution of the telescope~\cite{Wu:2014hta,Galli:2014kla,Allison:2015qca,Baumann:2017gkg}. For a single effective frequency channel with beam full width at half maximum (FWHM) $\theta_b$ and map sensitivity $\Delta X$---where $\Delta T$ and $\Delta P$ denote temperature and polarization sensitivities respectively---the noise power spectrum takes a Gaussian beam form:
\begin{equation}
    N_\ell^X = (\Delta X)^2 \exp\!\left\{ \frac{\ell(\ell+1)\,\theta_b^2}{8\ln 2} \right\}.
\end{equation}
This expression reflects the exponential suppression of sensitivity on angular scales finer than the beam. 

Next-generation CMB experiments are parameterized compactly in terms of a single effective frequency channel, described by three quantities: the noise level $\Delta T$, the beam width $\theta_b$, and the sky fraction $f_{\rm sky}$. Following the CMB-S4 Science Book~\cite{CMB-S4:2016ple} as our demonstrative baseline, the adopted values are $\theta_b = 2'$, $\Delta T = 1\,\mu{\rm K\,arcmin}$, and $f_{\rm sky} = 0.4$.

For the representative setup, the high-$\ell$ likelihood uses unlensed temperature and polarization spectra over the range $\ell_{\min} = 30$, $\ell_{\max}^T = 3000$, and $\ell_{\max}^P = 5000$.
The Fisher matrix for the CMB configuration is calculated directly over these multipole ranges.

For a representative future CMB mission like CMB-S4 experiment, the information is contained in the angular power spectra $C_\ell^X$ (where $X \in \{TT, TE, EE\}$). The CMB Fisher matrix is~\cite{CMB-S4:2016ple,Wu:2014hta,Galli:2014kla,Allison:2015qca,Baumann:2017gkg}:
\begin{equation}
    F_{\alpha\beta}^{\rm CMB} = \sum_{X,Y} \sum_{\ell=\ell_{\min}}^{\ell_{\max}} \frac{\partial C_\ell^X}{\partial \theta_\alpha} \left[\mathbf{C}_\ell^{XY}\right]^{-1} \frac{\partial C_\ell^Y}{\partial \theta_\beta},
\end{equation}
where $\mathbf{C}_\ell^{XY}$ is the analytic covariance matrix incorporating both cosmic variance and the representative CMB-S4 like instrumental noise spectra. Specifically, for each multipole $\ell$ and combinations $X=ab$, $Y=cd$ (with $a,b,c,d \in \{T,E\}$), the elements of this covariance matrix are given by:
\begin{equation}
\mathbf{C}_\ell^{(ab)(cd)} = \frac{1}{(2\ell + 1)f_{\mathrm{sky}}} \left[ \left(C_\ell^{ac} + N_\ell^{ac}\right)\left(C_\ell^{bd} + N_\ell^{bd}\right) + \left(C_\ell^{ad} + N_\ell^{ad}\right)\left(C_\ell^{bc} + N_\ell^{bc}\right) \right],
\end{equation}
where $C_\ell^{xy}$ are the theoretical CMB power spectra, $N_\ell^{xy}$ are the (Gaussian) noise spectra for CMB-S4 like representative mission, and $f_{\mathrm{sky}}$ is the effective sky fraction used in the cosmological analysis.

\section{All Parameter Fisher Contours}
\label{app:9params}
In this appendix, we provide the comprehensive results obtained from our joint Fisher analysis, expanding the parameter space to the full nine-parameter cosmological model. Empowered by the observational synergy between next-generation CMB surveys and LSS experiments, this extended analysis enables us to assess the constraining power of representative future CMB mission like CMB-S4, DESI-like, and radio telescope configurations (SKA1-Mid and PUMA) across all parameters simultaneously.

To illustrate the sensitivity of these future surveys to varying interaction strengths, we perform this full analysis for two distinct interaction regimes: SI$_\nu$ mode with coupling strength $\log_{10}G_{\rm eff} = -1.77$ and MI$_\nu$ mode with a fiducial coupling of $\log_{10}G_{\rm eff} = -5$. Fiducial values of other cosmological parameters are fixed as mentioned in Table.~\ref{tab:fiducial_params}. 

\begin{figure}[ht!]
\centering
\includegraphics[width=1.0\textwidth]{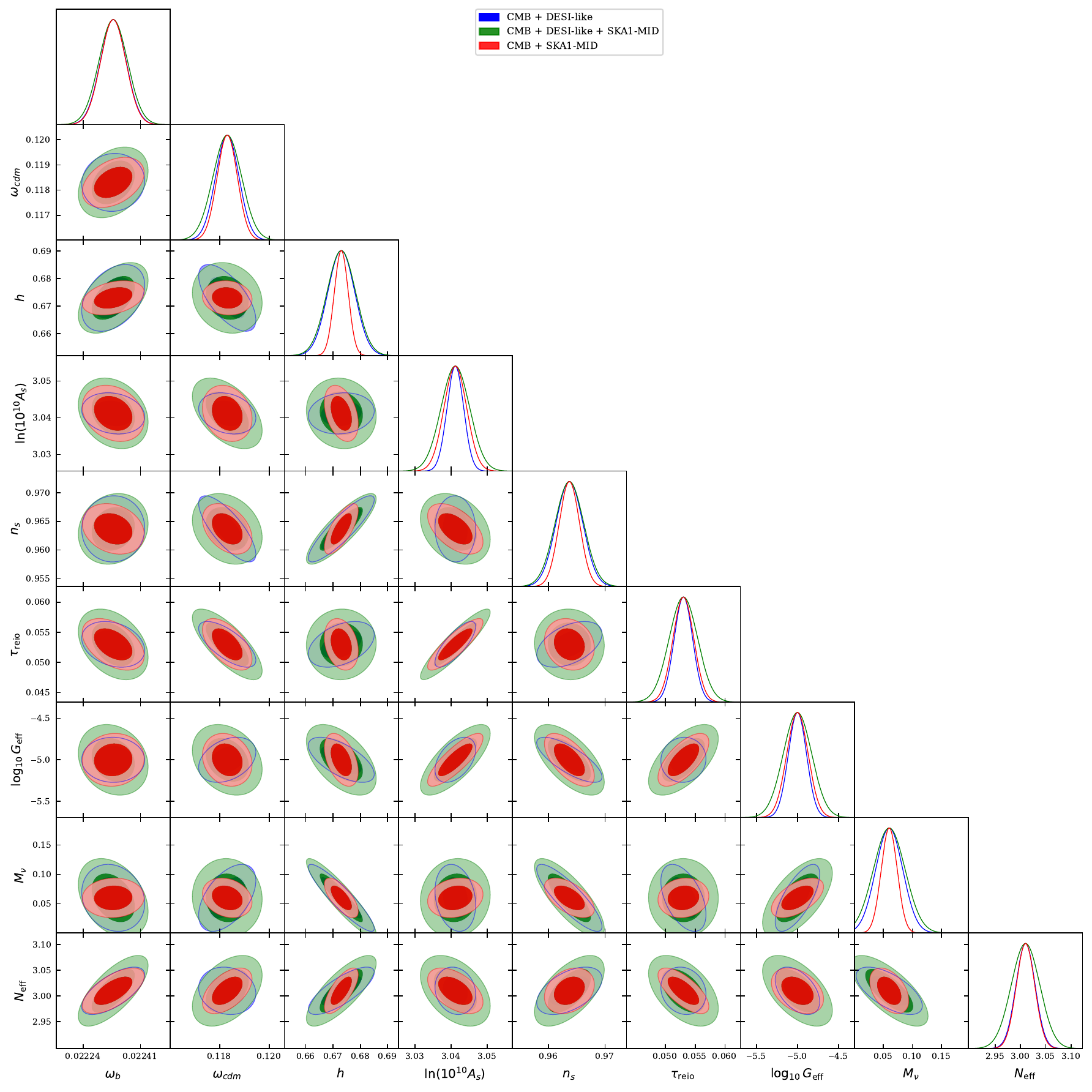}
\caption{Full nine-parameter marginalized $1\sigma$--$2\sigma$ Fisher contours in the MI$_\nu$ regime ($\log_{10}G_{\rm eff} = -5$) for CMB combined with DESI-like (blue), SKA1-Mid (red), and their joint combination (green). The $\log_{10}G_{\rm eff}$--$M_\nu$ degeneracy visible in the six-parameter analysis persists here, and extending to the full parameter space introduces additional correlations with $\tau_{\rm reio}$ and $\omega_{\rm cdm}$ that are absent in the compressed analysis.}
\label{fig:9params_MI_SKA}
\end{figure}

\begin{figure}[ht!]
\centering
\includegraphics[width=1.0\textwidth]{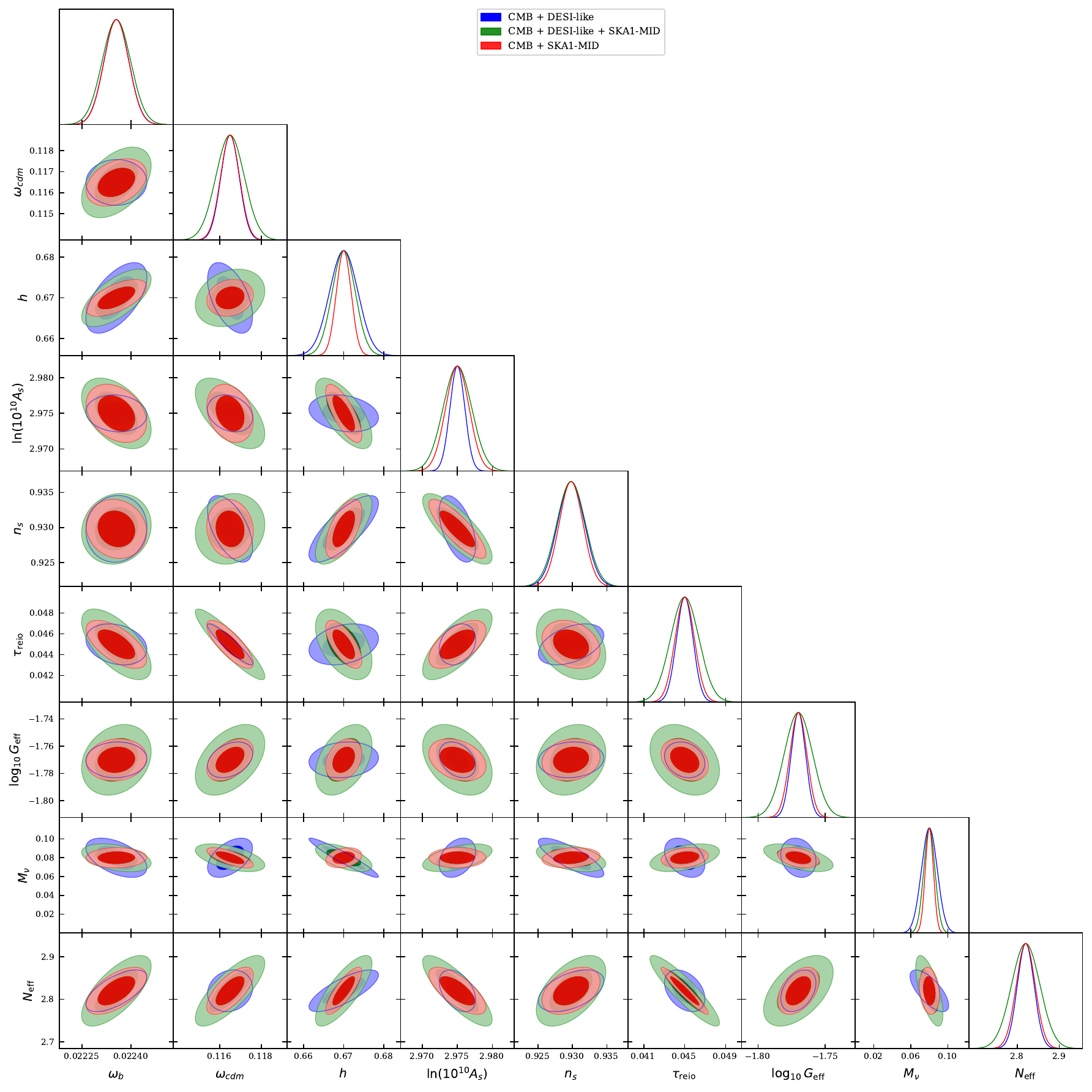}
\caption{Full nine-parameter marginalized $1\sigma$--$2\sigma$ Fisher contours in the SI$_\nu$ regime ($\log_{10}G_{\rm eff} = -1.77$) for CMB combined with DESI-like (blue), SKA1-Mid (red), and their joint combination (green). The contours are substantially tighter than the MI$_\nu$ counterpart across all parameters, consistent with the stronger signal amplitude in this regime.}
\label{fig:9params_SI_SKA}
\end{figure}

\begin{figure}[ht!]
\centering
\includegraphics[width=1.0\textwidth]{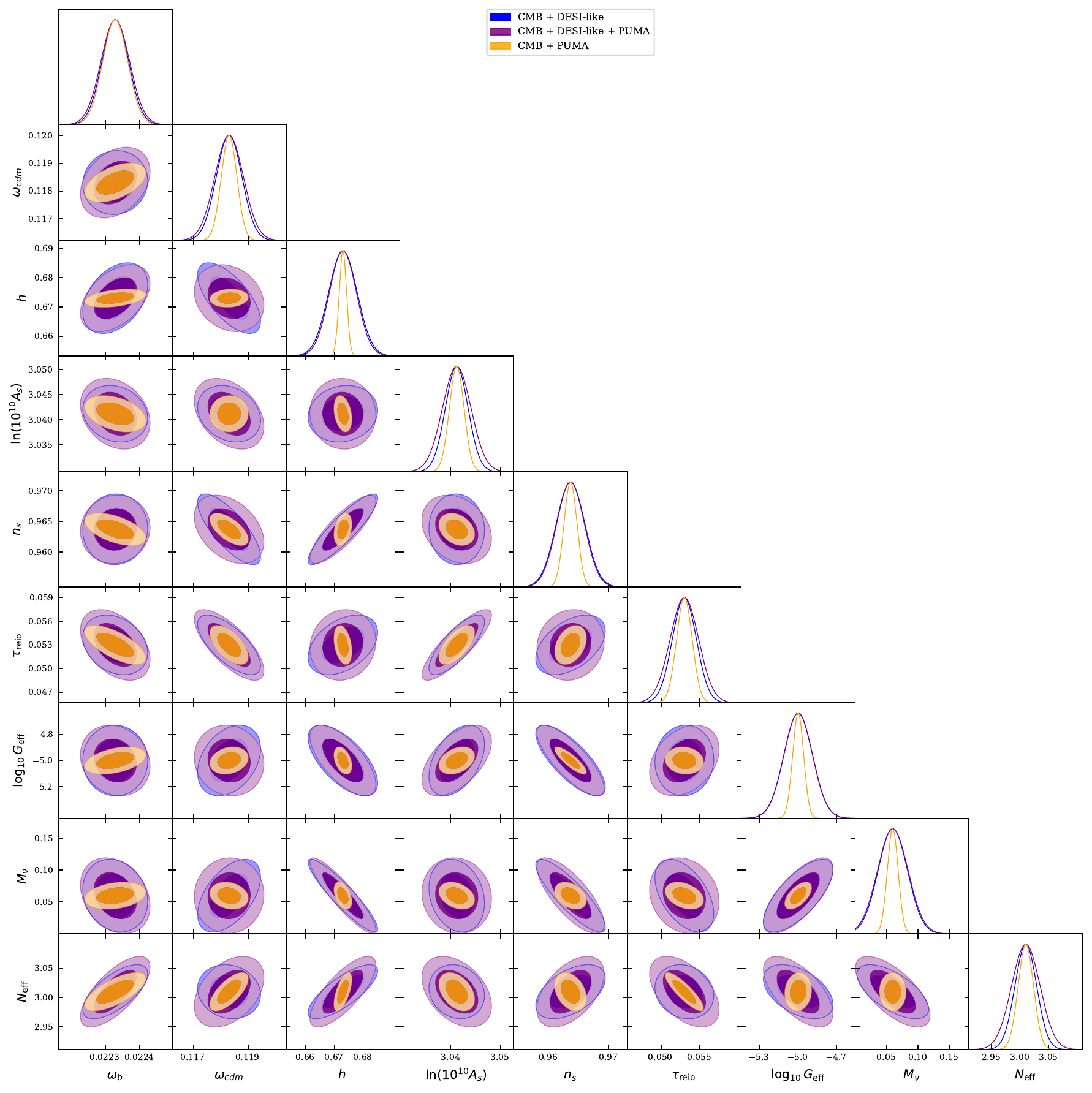}
\caption{ Full nine-parameter marginalized $1\sigma$--$2\sigma$ Fisher contours in the MI$_\nu$ regime ($\log_{10}G_{\rm eff} = -5$) for CMB-S4 combined with DESI-like (blue), PUMA (orange), and their joint combination (purple). The improvement over SKA1-Mid is most pronounced along $\log_{10}G_{\rm eff}$ and $h$, while the correlations involving $\omega_b$ and $\tau_{\rm reio}$ remain largely unchanged between the two radio configurations.}
\label{fig:9params_MI_PUMA}
\end{figure}

\begin{figure}[ht!]
\centering
\includegraphics[width=1.0\textwidth]{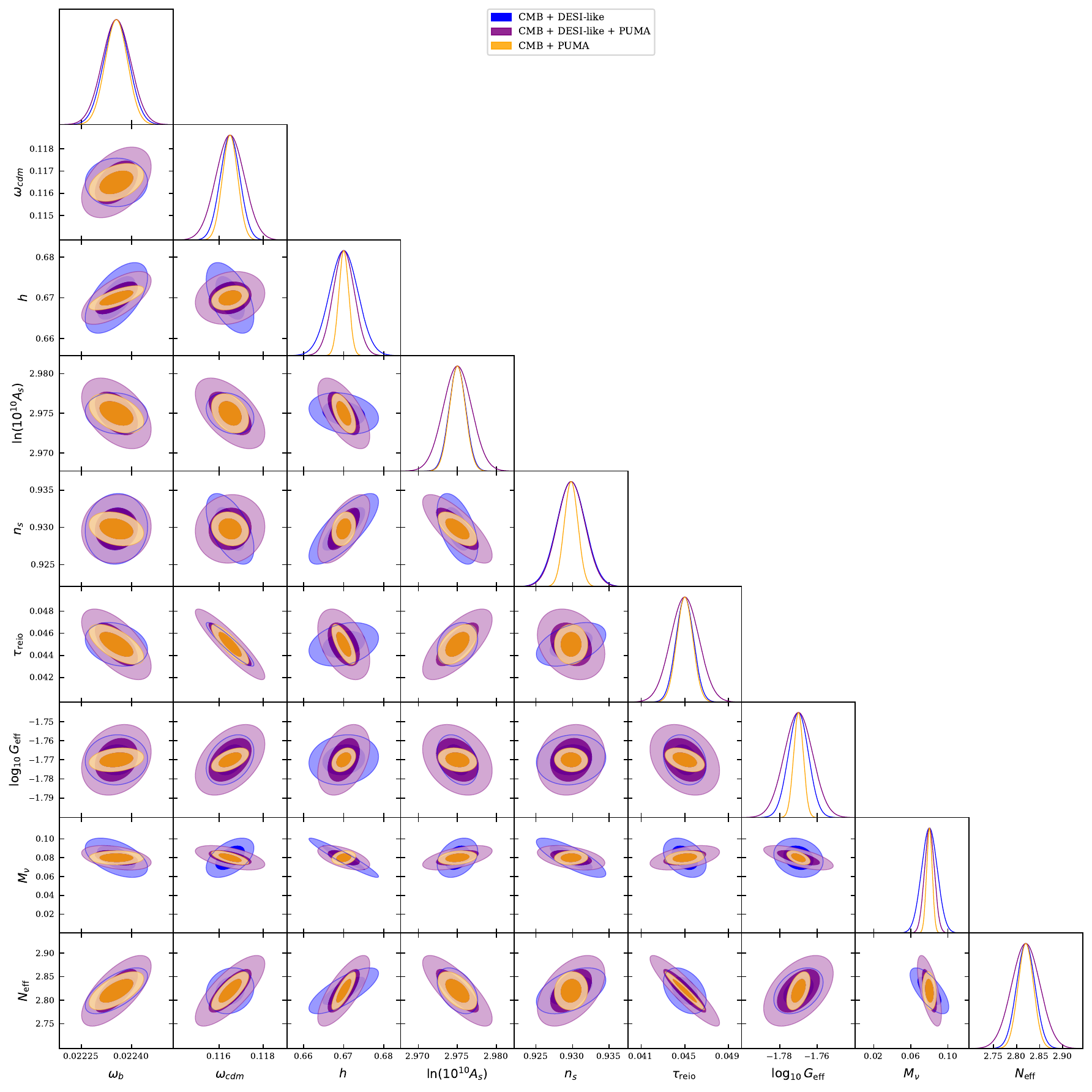}
\caption{Full nine-parameter marginalized $1\sigma$--$2\sigma$ Fisher contours in the SI$_\nu$ regime ($\log_{10}G_{\rm eff} = -1.77$) for CMB combined with DESI-like (blue), PUMA (orange), and their joint combination (purple). PUMA delivers the tightest posteriors across all nine parameters in the SI$_\nu$ regime, with particularly improved contours in the $\log_{10}G_{\rm eff}$--$M_\nu$ and $\ln(10^{10}A_s)$--$n_s$ planes.}
\label{fig:9params_SI_PUMA}
\end{figure}

\bibliographystyle{JHEP.bst}
\bibliography{refs.bib}

\end{document}